\documentclass[aps,prl,floats, twocolumn,superscriptaddress]{revtex4}
\usepackage[svgnames]{xcolor} 
\usepackage{graphicx,epsfig}
\usepackage{times} 
\usepackage{amssymb,amsmath,multirow,rotate,color}
\usepackage{xcolor}
\usepackage[normalem]{ulem}

\definecolor{albicocca}{rgb}{0.98, 0.7, 0.2}
\definecolor{internationalorange}{rgb}{1.0, 0.31, 0.0}
\definecolor{giocolor}{RGB}{0, 150, 100}

\definecolor{blueresponse}{RGB}{61, 133, 198}

\definecolor{qcblue}{RGB}{0, 61, 165}

\usepackage{float}
\usepackage{lipsum} 
\usepackage{ragged2e}
\usepackage{soul}
\DeclareMathAlphabet\mathbfcal{OMS}{cmsy}{b}{n}

\usepackage{booktabs}
\usepackage{orcidlink}

\begin{document}

\title{Group dynamics shape contagion onsets and multistable active phases\\ under collective reinforcement}

\author{Santiago Lamata-Otín\,\orcidlink{0009-0004-0247-4792}}
\affiliation{GOTHAM lab, Institute of Biocomputation and Physics of
Complex Systems (BIFI), University of Zaragoza, 50018 Zaragoza, Spain}
\affiliation{Department of Condensed Matter Physics, University of Zaragoza, 50009 Zaragoza, Spain}
\affiliation{
Vermont Complex Systems Institute, University of Vermont, Burlington, VT 05405 USA}

\author{Federico Malizia\,\orcidlink{0000-0003-0991-3861}}
\affiliation{Department of Network and Data Science, Central European University, Vienna, Austria}

\author{Leah A. Keating\,\orcidlink{0000-0001-6911-7432}}
\affiliation{
Vermont Complex Systems Institute, University of Vermont, Burlington, VT 05405 USA}
\affiliation{Department of Computer Science, University of Vermont, Burlington, VT 05405 USA}

\author{Guillaume St-Onge\,\orcidlink{0000-0001-5415-3748}}
\affiliation{Laboratory for the Modeling of Biological and Socio-technical Systems, Northeastern University, Boston MA, USA}
\affiliation{The Roux Institute, Northeastern University, Portland ME, USA}

\author{Vito Latora\,\orcidlink{0000-0002-0984-8038}}
\affiliation{Department of Physics and Astronomy,  University of Catania, 95125 Catania, Italy}
\affiliation{School of Mathematical Sciences, Queen Mary University of London, London E1 4NS, United Kingdom}
\affiliation{Complexity Science Hub Vienna, A-1080 Vienna, Austria}

\author{Jesús Gómez-Gardeñes\,\orcidlink{0000-0001-5204-1937}}
\email{gardenes@unizar.es}
\thanks{Corresponding author}
\affiliation{GOTHAM lab, Institute of Biocomputation and Physics of
Complex Systems (BIFI), University of Zaragoza, 50018 Zaragoza, Spain}
\affiliation{Department of Condensed Matter Physics, University of Zaragoza, 50009 Zaragoza, Spain}

\author{Laurent Hébert-Dufresne\,\orcidlink{0000-0002-0008-3673}}
\email{laurent.hebert-dufresne@uvm.edu}
\thanks{Corresponding author}
\affiliation{
Vermont Complex Systems Institute, University of Vermont, Burlington, VT 05405 USA}
\affiliation{Department of Computer Science, University of Vermont, Burlington, VT 05405 USA}
\affiliation{Complexity Science Hub Vienna, A-1080 Vienna, Austria}
\affiliation{Santa Fe Institute, Santa Fe, NM 87501, USA}

\date{\today}


\begin{abstract}
Group-based reinforcement can induce discontinuous transitions from inactive to active phases in higher-order contagion models. However, these results are typically obtained on static interaction structures or within mean-field approximations that neglect temporal changes in group composition. Here, we show that group dynamics is not a secondary effect but a central aspect that determines the macroscopic transition class of higher-order contagion processes. We develop an analytically tractable approximate master equation model that effectively interpolates between quenched and mean-field limits through a group composition turnover rate. Our results reveal the rich impact of time-varying structures: it can induce discontinuous phase transition, broaden the bistable region, and at the same time promote or suppress contagion near criticality. Moreover, when real-world turnover rates and group-size heterogeneity are taken into account, the system exhibits a qualitatively richer phase diagram with four distinct dynamical phases, combining continuous or discontinuous transitions with localized or delocalized activity.In localized regimes, we uncover multistable active phases with multiple coexisting active states, which are observed in neither the annealed nor the quenched limits, and extend classical absorbing-active bistability. Finally, we demonstrate that the emergence of discontinuous transitions in real-world systems requires stronger nonlinear reinforcement than previously thought, indicating that simulations in static structures can yield qualitatively misleading predictions.
\end{abstract}

\maketitle

In social systems, the adoption of new behaviors or ideas often requires more than a single exposure: individuals may need reinforcement from multiple peers before changing state \cite{centola2007complex,centola2010spread,guilbeault2018complex}. This feature has motivated a broad class of complex contagion models \cite{granovetter1978threshold,watts2002simple,dodds2005generalized,centola2007cascade,o2015mathematical,guilbeault2021topological,liu1987dynamical,assis2009discontinuous,gomez2016explosive} traditionally based on threshold rules \cite{granovetter1978threshold,watts2002simple,dodds2005generalized,centola2007cascade,o2015mathematical,guilbeault2021topological} or on nonlinear synergistic mechanisms \cite{liu1987dynamical,assis2009discontinuous,gomez2016explosive}. 
Recent work by Iacopini et al. \cite{iacopini2019simplicial} revitalized this field by representing reinforcement in terms of interactions in groups of two or more 
individuals. This formulation of complex contagion showed that, even when operating solely within higher-order interactions rather than across entire neighborhoods, group reinforcement can change the nature of the transition from a no-adopter to an active adoption state, from continuous to discontinuous  \cite{dodds2005generalized,gomez2016explosive}. This result triggered the study of diverse dynamical processes on systems with higher-order interactions \cite{landry2020effect,st2022influential,lamata2024integrating,meloni2025higher,keating2025loops,ferraz2024contagion,wang2024epidemic,sun2021higher,sun2023dynamic,di2024percolation,tanaka2011multistable,millan2020explosive,skardal2020higher,gambuzza2021stability,zhang2024deeper,lamata2025hyperedge,perez2025social,burgio2025characteristic,gomez2011evolutionary,gomez2011disentangling,alvarez2021,civilini2023explosive,meng2025promoting,wang2026strategy,breton2025explosive,millan2025topology,battiston2025higher,battiston2026collective}, including contagion \cite{landry2020effect,st2022influential,lamata2024integrating,meloni2025higher,keating2025loops,ferraz2024contagion,wang2024epidemic}, percolation \cite{sun2021higher,sun2023dynamic,di2024percolation}, synchronization \cite{tanaka2011multistable,millan2020explosive,skardal2020higher,gambuzza2021stability,zhang2024deeper,lamata2025hyperedge}, social polarization \cite{perez2025social}, self-organization \cite{burgio2025characteristic}, and evolutionary dynamics \cite{gomez2011evolutionary,gomez2011disentangling,alvarez2021,civilini2023explosive,meng2025promoting,wang2026strategy}. 
\medskip

Despite this recent burst in the study of higher-order dynamics, most of the works neglect that real group interactions are inherently temporal \cite{palla2007quantifying}. This implies that individuals can join and leave groups on time scales comparable to those relevant to the dynamical process under study. In particular, for higher-order contagion processes, two regimes (capturing oposing scenarios) are usually considered. On one hand, those studies capturing discontinuous contagion transitions typically adopt the annealed (mean-field) limit \cite{iacopini2019simplicial}, corresponding to infinitely fast reshuffling. Under this homogeneous mean-field approximation, the invasion threshold, {\em i.e.}, the critical condition under which a small seed of adopters can successfully spread, is independent of the nonlinearity of the contagion dynamics \cite{iacopini2019simplicial}. On  the other hand, in the opposite limit of a static (quenched) network structure, the threshold can even diverge as the network becomes fragmented \cite{hebert2010propagation}. 
\medskip

The few efforts devoted to understanding contagion dynamics on time-varying structures report apparently conflicting trends \cite{valdano2015analytical,st2018phase,chowdhary2021simplicial}. In particular, classical results for time-varying graphs \cite{valdano2015analytical,st2018phase} indicate that temporal variability in pairwise interactions enhances mixing,
thus lowering the invasion threshold compared to static counterparts. In contrast, when group interactions are considered, \cite{chowdhary2021simplicial} shows that the invasion threshold shifts towards larger values in time-varying higher-order structures, even in the case of linear higher-order contagion. These contrasting findings raise the question of whether group-based contagion processes are intrinsically different from pairwise ones, or whether the discrepancy originates from the specific manner in which temporality is modeled. Furthermore, beyond the effect of temporal variability on critical thresholds, its impact on the properties of the active (adoption) phase remains largely unexplored.
\medskip

In this work, we develop an analytically tractable approximate master equation (AME) model that couples higher-order contagion dynamics with temporal variability in the composition of groups. Our framework leads to three important results. First, we find that increasing group switching rate can qualitatively alter the nature of the phase transition, transforming a continuous transition into a discontinuous one, and enlarging the bistable region in systems with a homogeneous group-size distribution. In this setting, we derive that the invasion threshold becomes non-monotonic in the group switching rate, implying the existence of an optimal group switching rate beyond which temporality suppresses adoption. Second, when we  introduce group-size heterogeneity as that  observed in real-world systems, we find that the phase portrait is characterized by four distinct dynamical regimes, emerging from the interplay between nonlinear reinforcement and mesoscopic localization. In particular, we identify a region exhibiting a hybrid continuous transition with active bistability, as well as a region featuring a discontinuous hybrid transition with three coexisting stable states: the absorbing state and two active states.
To round off, our third main finding is that the empirically measured temporal and structural features of real-world social systems require extremely strong nonlinear reinforcement to produce discontinuous transitions. This last result indicates that static or mean-field representations may yield qualitatively misleading predictions.
\medskip

\subsection{Model of complex contagion on temporal higher-order networks}

\begin{figure}
    \centering
    \includegraphics[width=1\linewidth]{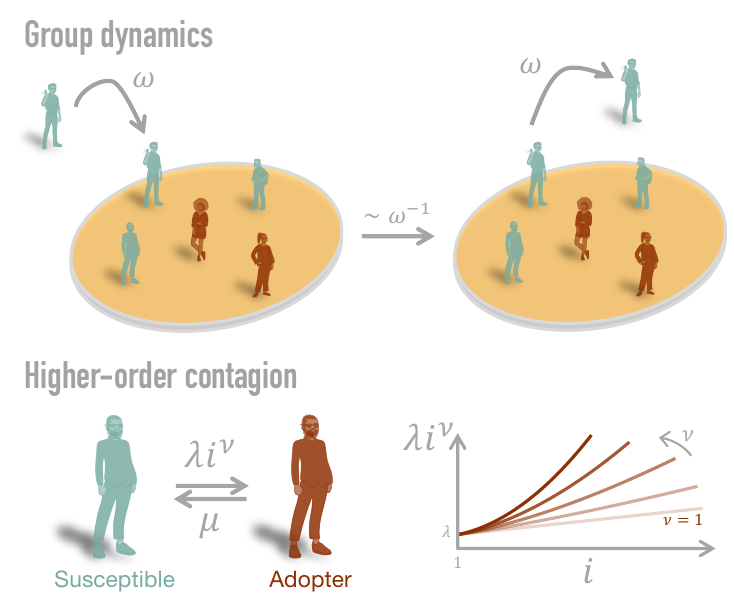}
    \caption{\textbf{Temporal higher-order contagion model.} 
In our model,  a susceptible individual in a group of size $n$   with other $i$ adopters becomes infected at rate $\beta(n,i)=\lambda i^\nu$. The synergy exponent $\nu$ controls the group  reinforcement.  Adopters recover at rate $\mu$. While contagion unfolds within each group, the individuals move between groups at a  
at rate $\omega$.}
    \label{fig:Fig1}
\end{figure}

We consider a set of groups whose size $n$ follows a distribution $\{p_n\}$, and a population of individuals whose number of group memberships (i.e., the number of groups an individual participates in) $k$ follows a distribution $\{g_k\}$. In our model of complex contagions coupled with a time varying group structure, two parallel processes govern the dynamics of the system (see Fig. \ref{fig:Fig1}). \textit{(i) Higher-order contagion}: a susceptible individual in a group of size $n$ with $i$ adopters becomes an adopter at rate $\beta(n,i)=\lambda i^{\nu}$, where $\lambda$ is the intrinsic adoption rate and the synergy exponent $\nu$ controls the strength of the reinforcement; moreover, adopters revert to the susceptible state at rate $\mu$. \cite{st2021universal,st2022influential}. \textit{(ii) Group dynamics}: Random individuals are swapped between random groups at a  rate $\omega$. In this way, all changes in group composition are regulated by a single node-level group switching process controlled by the parameter $\omega$.
\medskip

The resulting coevolving  dynamics of complex contagion and temporal higher-order network are described by a set of AMEs, the Eqs. (\ref{eq:sm})-(\ref{eq:r}) in Methods, that we name $\omega$AMEs. The model tracks the density $s_k$ of susceptible individuals with membership $k$,  and the fraction of groups of size $n$ containing $i$ adopters, $f_{n,i}$. From these quantities, we compute the stationary fraction of the adopter population $I^\star$ (see Eq. (\ref{eq:I(t)}) in Methods), which we use as order parameter throughout the following sections to characterize the macroscopic behavior. We refer to Eqs. (\ref{eq:M})-(\ref{eq:stationary}) in Methods and Supplementary Note 1 for details on how the model is solved.
\medskip

\begin{figure*}
    \centering
    \includegraphics[width=1\linewidth]{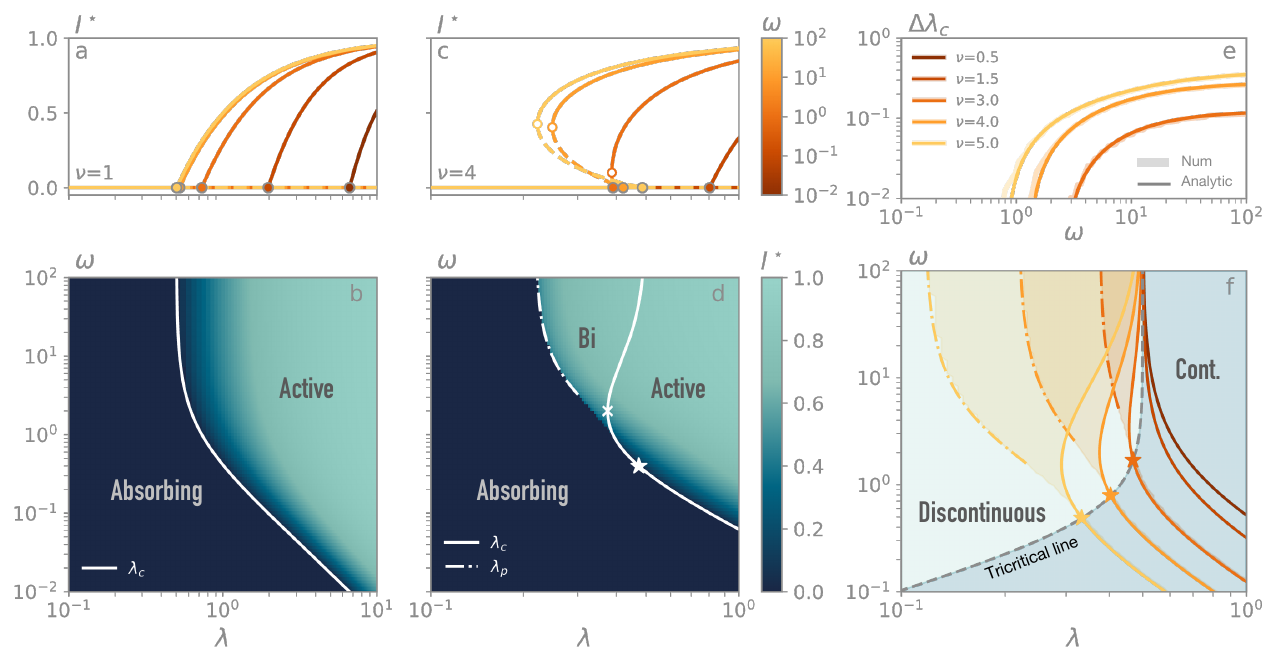}
    \caption{\textbf{Group dynamics reshapes the critical behavior of higher-order contagion.} 
    \textbf{a–b} Stationary prevalence $I^\star$ and corresponding phase diagram for linear contagion ($\nu=1$) as a function of the spreading rate $\lambda$ and of the group switching rate $\omega$. Increasing the rate at which individuals switch groups lowers the invasion threshold, and the transition remains continuous.  
    \textbf{c–d} Same as in previous panels for nonlinear contagion ($\nu=4$). Here, group dynamics  fundamentally alters the transition: the invasion threshold becomes non-monotonic in $\omega$, and sufficiently strong synergy produces a discontinuous transition. Analytical predictions in Eqs. (\ref{eq:inv_thres_compact}) and (\ref{eq:lambda-bi-final}) for the invasion threshold $\lambda_c(\omega)$ and the persistence threshold $\lambda_p(\omega)$ are shown throughout, the tricritical point derived using Eq. (\ref{eq:bi_threshold}) in Methods is marked with a star in panel~d, and the minimum of the invasion threshold according to Eqs. (\ref{eq:wstar_general})-(\ref{eq:lamb_star_ge}) is marked with a cross in panel~d.  
    \textbf{e} Width of the bistable region, $\Delta\lambda(\omega)=\lambda_c(\omega)-\lambda_p(\omega)$, as a function of the group switching rate for different values of the synergy exponent $\nu$.
    \textbf{f} Bifurcation diagram showing the analytically derived tricritical line separating continuous and discontinuous transitions, together with the corresponding invasion and persistence thresholds for the nonlinearities considered in panel~e. All analytical curves closely match the steady-state solutions of the AMEs. In all panels, each individual belongs to one group only ($g_k=\delta_{k,1}$), the groups have homogeneous size ($p_n=\delta_{n,3}$), and $\mu=1$. 
    }

    \label{fig:Fig2}
\end{figure*}

\subsection{Group dynamics reshapes the phase transition of higher-order contagion} 

Temporal variations in the group composition have fundamental effects  on the phase transition of a higher-order contagion process. In Fig.~\ref{fig:Fig2} 
we study the case when individuals can  participate in only one face-to-face interaction at a time ($g_k=\delta_{k,1}$).
In the quenched limit, $\omega=0$, group composition is frozen, and individuals remain confined to their initial interaction patterns which globally consist of a set of disjoint groups. Under these conditions, contagion cannot invade: adopters are unable to sustain activity due to the finite size of the groups and the absence of coupling between them, leading the system to collapse into the absorbing state $I^\star=0$ \cite{hebert2010propagation}. The invasion threshold $\lambda_c$, i.e., the critical condition under which a small seed of adopters can successfully spread, therefore diverges (see Supplementary Notes 1, 2 and 3 for the derivation): 
\begin{eqnarray}
\lim_{\omega \to 0} \lambda_c(\omega) = \infty,
\label{eq:quenched}
\end{eqnarray}
capturing the fact that there are no adoption transitions in finite populations; or in the case of our model, no phase transitions in infinite populations made of disconnected finite groups.
\medskip

When $\omega>0$, the temporal evolution in the composition of groups changes this picture qualitatively. Even low group switching rates expose individuals to multiple group configurations over time. As a result, a transition from the absorbing to the active phase emerges. In the limit  
$\omega\rightarrow\infty$ of 
infinitely fast reshuffling, individuals effectively sample all possible group realizations, yielding the mean-field critical point
\begin{equation}
\lim_{\omega \to \infty} \lambda_c(\omega)
=
\frac{\mu\, \langle n\rangle}{\langle k\rangle\, \langle n(n-1)\rangle},
\label{eq:mean_field}
\end{equation}
where $\langle n\rangle$ and $\langle k\rangle$ are the mean group size and membership respectively. Eq. (\ref{eq:mean_field}) holds for both linear and nonlinear contagion (see Supplementary Note 3 for the derivation). The full curve interpolating between these limits matches the steady-state $\omega$AME solutions in Fig.~\ref{fig:Fig2}a–d. This general solution is reported in Eq. (\ref{eq:inv_thres}) of Methods and its derivation is reported in the Supplementary Note 2. 
\medskip

Having established that group dynamics induces a transition absent in the quenched limit, i.e., in the absence of shuffling, we next examine how collective reinforcement modifies the nature of this transition. From Fig.~\ref{fig:Fig2}.a-b it is clear that contagion is linear ($\nu=1$), the transition is always continuous: $I^\star$ grows smoothly from zero as $\lambda$ crosses $\lambda_c$. However, in the presence of nonlinear reinforcement ($\nu>1$), temporality can fundamentally alter the order of the phase transition. As shown in Fig.~\ref{fig:Fig2}.c-d, for sufficiently large values of the synergy exponent,  increasing the group switching rate causes the continuous transition to become discontinuous and introduces bistability, where both the absorbing and active states are locally stable. The tricritical condition, which separates the continuous transition regime from the bistable discontinuous transition regime (marked with a star symbol in Fig.~\ref{fig:Fig2}.d), is reported in Eq. (\ref{eq:bi_threshold}) in Methods and derived in the Supplementary Note 2. As shown in Fig.~\ref{fig:Fig2}.e-f, the width of the bistable region increases with both the group switching rate $\omega$ and the synergy exponent $\nu$, revealing that temporal group switching enhances the reinforcement effects needed to generate explosive adoption events \cite{d2019explosive}.
\medskip

An even more important consequence of group dynamics is that, when collective reinforcement is strong  ($\nu>1$), the adoption threshold $\lambda_c(\omega)$ becomes non-monotonic in the group switching rate. Figures~\ref{fig:Fig2}c–d show that $\lambda_c(\omega)$ initially decreases as the group switching rate increases, reaching a finite minimum value at $\omega^\star$. This indicates that temporality enhances contagion only up to an optimal mixing rate; beyond this point, additional reshuffling suppresses the reinforcement required to propagate adoption, and $\lambda_c(\omega)$ increases again. The general expression for the invasion threshold in systems with only three-body interactions ($p_n=\delta_{n,3}$) is derived in Eq.~(\ref{eq:lambda_general_k}) and Supplementary Note 3. For individuals participating in only one face-to-face interaction at a time ($g_k=\delta_{k,1}$), it reduces to
\begin{equation}
    \lambda_{c,n=3}(\omega)
=
\frac{\mu+\omega}{2^{\nu}}
\left[
\sqrt{1 + \frac{2^{\nu}\mu}{\omega}}
- 1
\right].
\label{eq:inv_thres_compact}
\end{equation} 
This expression has a finite minimum when $\nu>2$, corresponding to an optimal group switching rate (see Eqs.~(\ref{eq:wstar_general})–(\ref{eq:lamb_star_ge})). 
It is important to highlight that the non-monotonicity of the adoption threshold is characteristic of group interactions in homogeneous systems. In fact, when only pairwise interactions are present ($p_n=\delta_{n,2}$), the critical line reduces to $\lambda_{c,n=2}(\omega)=\mu+\mu^2/\omega$,  which is strictly monotonic in $\omega$ 
and independent of~$\nu$.
\medskip

Allowing agents to participate in multiple groups in parallel ($g_k=\delta_{k,\kappa},\; \kappa>1$) defines the transition even in the absence of group temporal changes and shifts the minimum of the invasion threshold towards zero, potentially pushing it outside the physical domain. As a result, $\lambda_c(\omega)$ becomes strictly increasing in $\omega$. In this regime, increasing the group switching rate consistently suppresses adoption, in contrast with linear contagion dynamics, where group dynamics facilitates spreading. Larger typical group sizes have a similar effect in the invasion threshold, as they amplify collective reinforcement (see Supplementary Fig. 2 in Supplementary Note 4 for further details).
\medskip

In contrast, the persistence threshold, i.e., the critical condition below which an established active state cannot be sustained, remains strictly monotonic in $\omega$. This occurs because, when starting from a fully adopter population, groups contain a sufficient density of adopters such that temporal reshuffling does not disrupt the reinforcement needed to sustain adoption. Instead, group switching primarily increases contacts between susceptible and adopter individuals. For large group switching rates and $p_n=\delta_{n,3}$, analytical expressions for the persistence threshold $\lambda_p(\omega)$ and the width of the bistable region $\Delta\lambda(\omega)=\lambda_c(\omega)-\lambda_p(\omega)$ are derived in Methods and Supplementary Note 5, and match the stationary solutions of the $\omega$AMEs (Fig.~\ref{fig:Fig2}).

\begin{figure*}
    \centering
    \includegraphics[width=1\linewidth]{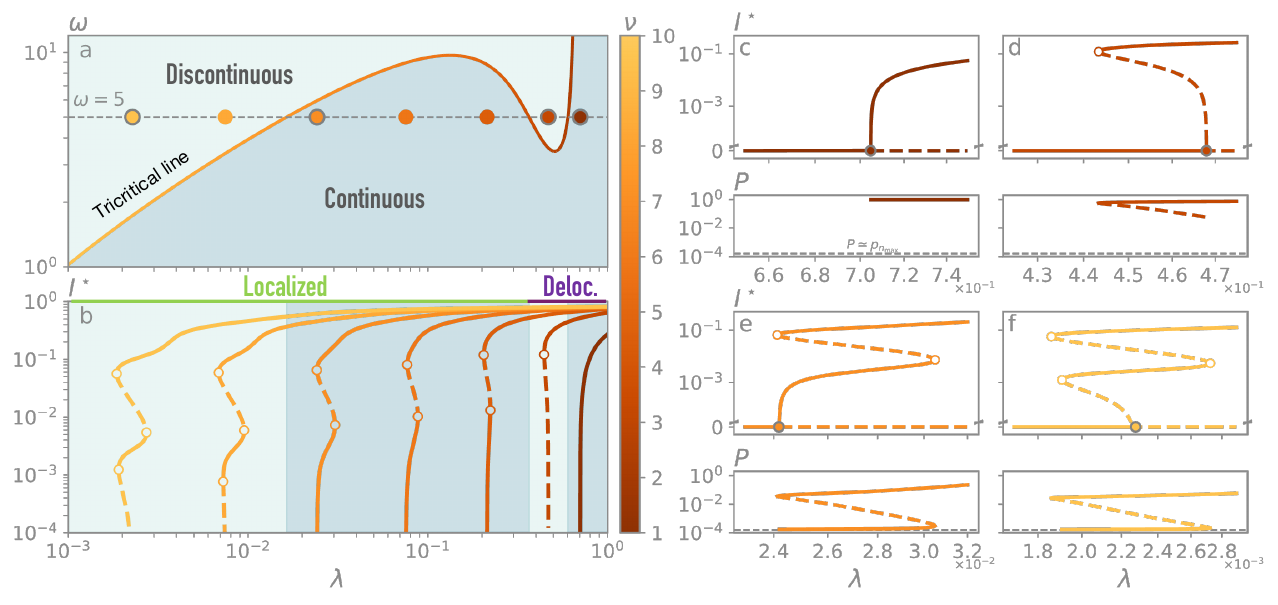}
    \caption{\textbf{Interplay between group dynamics and heterogeneity yields multistability.} Using the empirically observed group-size and membership distributions of a real-world system, namely the high school dataset of Ref. \cite{mastrandrea2015contact}, the panels show: \textbf{a} Tricritical line in the $(\omega,\lambda)$ plane with the color corresponding to the value of the synergy exponent $\nu$. \textbf{b} Stationary prevalence $I^\star$ as a function of the adoption rate $\lambda$ for selected values of the synergy exponent $\{\nu\}=\{1, 3.2,4.5,5.75,7,8.25,9.5\}$ and $\omega=5$. Solid (dashed) lines denote stable (unstable) stationary states. \textbf{c}-\textbf{f} Representative phase diagrams illustrating four distinct dynamical regimes: continuous transition in panel \textbf{c} for $\nu=1$, discontinuous transition with absorbing–active bistability in panel \textbf{d} for $\nu=3.2$, continuous hybrid transition with active bistability in panel \textbf{e} for $\nu=7$, and discontinuous hybrid transition with three coexisting stable states in panel \textbf{f} for $\nu=9.5$. Bottom rows show the effective participation ratio $P$ quantifying the degree of localization of activity across groups. In all panels, $\mu=1$.
    }
    \label{fig:Fig3}
\end{figure*}

\subsection{Phase transitions and multistability induced by group dynamics and heterogeneity}

Having established how group dynamics 
reshapes the adoption onset of higher-order contagions in 
synthetic homogeneous structures, in Fig.~\ref{fig:Fig3} we now turn to real-world social systems from various contexts~\cite{genois2018can,isella2011s,vanhems2013estimating,mastrandrea2015contact,stehle2011high,ozella2021using,sapiezynski2019interaction}, where group sizes are heterogeneous. Each dataset is characterized by a group-size distribution $\{p_n\}$ and a membership distribution $\{g_k\}$, the latter being $g_k\approx\delta_{1,k}$ in all contexts (see Fig.~2 in Supplementary Note 6).
\medskip

These structural parameters determine the tricritical line $[\lambda_c^\star,\nu_c^\star](\omega)$ separating continuous and discontinuous transitions. For homogeneous group sizes, the tricritical line is a monotonic function of the group switching rate (Fig.~\ref{fig:Fig2}.f). However, once we incorporate the empirical size and membership distributions of a real system, the tricritical boundary becomes non-monotonic and develops a folded structure as shown in Fig.~\ref{fig:Fig3}a for the high-school dataset \cite{mastrandrea2015contact} 
(see Fig. 3a in Supplementary Note 7  for the other datasets). As a consequence, for a fixed group switching rate, the tricritical condition in Eq. (\ref{eq:bi_threshold}) can admit multiple solutions: for instance, at $\omega=5$ the horizontal cut intersects the tricritical curve three times.
\medskip

The dynamical consequences of the nontrivial folded  tricritical boundary are shown in Fig.~\ref{fig:Fig3}.b where we show the stationary prevalence as a function of the intrinsic adoption rate for seven values of the synergy exponent $\nu$, with the corresponding invasion thresholds indicated in Fig.~\ref{fig:Fig3}a. Continuous segments of the curves correspond to linearly stable stationary states, whereas dashed segments denote unstable solutions. As collective reinforcement ($\nu$) increases, the system undergoes four qualitatively distinct dynamical regimes, illustrated in Fig.~\ref{fig:Fig3}.c–f.

For weak reinforcement, $\nu=1$, the transition is continuous and resembles that of the classical SIS dynamics (Fig.~\ref{fig:Fig3}c). At intermediate $\nu$, reinforcement within groups becomes strong enough to induce a discontinuous transition with bistability between absorbing and active states (see Fig.~\ref{fig:Fig3}d for $\nu=3.2$). 
As the synergy exponent $\nu$ increases further, the qualitative structure of the phase diagram changes again. In Fig.~\ref{fig:Fig3}.e–f, the active phase no longer consists of a single endemic state. Instead, multiple active branches appear. For $\nu=7$, in Fig.~\ref{fig:Fig3}e we observe a hybrid continuous transition with active bistability: activity emerges continuously from zero, yet two stable active branches coexist over a finite parameter range. For even stronger synergy ($\nu=9.5$ in Fig.~\ref{fig:Fig3}f), the transition becomes a discontinuous hybrid one, with three coexisting stable states, namely the absorbing state and two distinct active states.

To understand the emergence of these active states, we introduce a measure, the effective participation ratio $P$, to quantify the effective number of groups contributing to adoption:
\begin{equation}
P=\frac{\left(\sum_n p_n I_n^{2}\right)^{2}}{\sum_n p_n I_n^{4}}\in[p_{n_{\text{max}}},1]\;,
\end{equation}
where $I_n=\sum_{i=0}^{n} i f_{n,i}/n$ denotes the stationary prevalence in groups of size $n$. This effective participation ratio becomes large when activity is broadly distributed across groups, corresponding to a delocalized state, whereas it remains small when adoption is concentrated in a limited subset, corresponding to a localized active state (see bottom rows in Fig.~\ref{fig:Fig3}c–f) \cite{st2022influential}.
While the active states in Fig.~\ref{fig:Fig3}c–d correspond to delocalized activity, the ones in Fig.~\ref{fig:Fig3}e–f exhibit reduced participation ratios, signaling mesoscopic localization. In a nutshell, the lower stable state in both panels is sustained by the adopters in the largest groups, whereas the upper state remains only partially localized.
\medskip 

In Supplementary Fig.~3b we further assess the role of group-size heterogeneity using synthetic structures with heterogeneous group sizes. Supplementary Fig.~4 complements this analysis by presenting the phase diagram analogous to Fig.~\ref{fig:Fig3}b in the quenched and annealed limits. 
The quenched regime displays continuous transitions with non-monotonic prevalence growth and mesoscale localization plateaus \cite{st2021social, st2021master,st2022influential}, whereas the annealed regime exhibits the discontinuous transitions predicted by mean-field analysis \cite{iacopini2019simplicial}. Importantly, multistable active phases with multiple coexisting active states are observed in neither limit, confirming that multistability is intrinsically driven by group dynamics and emerges only at finite switching rates.

\subsection{Temporal reshuffling determines the transition class in real-world systems}

\begin{figure}
    \centering
    \includegraphics[width=1\linewidth]{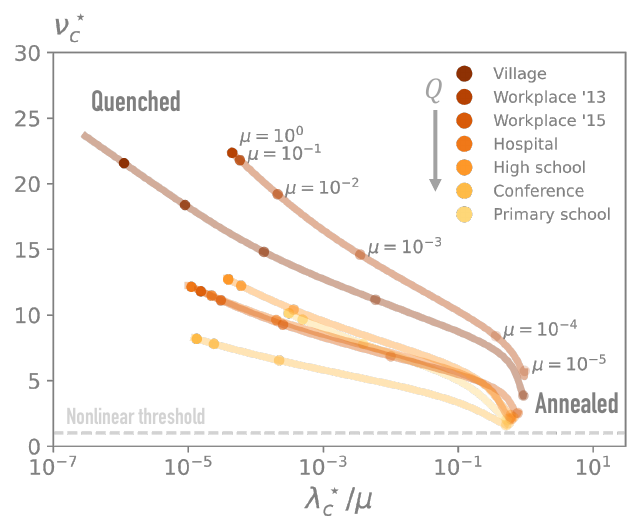}
    \caption{\textbf{Tricritical point across dynamical timescales in real-world systems.} For each dataset ~\cite{genois2018can,isella2011s,vanhems2013estimating,mastrandrea2015contact,stehle2011high,ozella2021using,sapiezynski2019interaction}, colored curves trace how the critical point shifts as the group switching 
    rate $\omega$ is varied from the 
    quenched to the annealed limit; markers indicate the tricritical points corresponding to 
    the empirically measured rate $\langle\omega\rangle$ for six different values of the contagion timescales, from $\mu = 10^{0}$ to $\mu = 10^{-5}$. Colors indicate the rank of the effective structural coupling $Q$ introduced in Eq.~(\ref{eq:Q_effective}).}
    \label{fig:fig4}
\end{figure}

As shown in the previous section, the structural distributions $\{p_n\}$ and $\{g_k\}$ define a tricritical line separating regions with continuous and discontinuous transitions as the group switching rate varies (see Fig. \ref{fig:Fig3}a). In Fig.~\ref{fig:fig4} we represent this tricritical line in the $(\nu, \lambda)$ space for each real-world system, illustrating how the tricritical point moves continuously from the quenched (static) to the annealed (mean-field) limit when varying the group switching rate. Across all the real-world systems investigated, discontinuous transitions occur only for consistently large synergy ($\nu_c^\star$), highlighting that strong collective reinforcement is necessary for bistability.
\medskip

However, locating the actual empirical systems along the tricritical line requires incorporating their intrinsic temporal scales. The transition class is not determined by structural heterogeneity alone, but by the competition 
between two time scales, 
the time scale of the contagion process ($\mu^{-1}$) and that 
of the motion of individuals between groups ($\omega^{-1}$).  
To quantify the latter, we introduce the effective group switching rate $\langle \omega\rangle$, defined as the inverse of the average residence time of individuals within groups (see Eq. (\ref{eq:eff_plasticity}) in Methods and Supplementary Note~6). 
\medskip

Fig. \ref{fig:fig4} shows the tricritical point corresponding to the empirical value $\langle\omega\rangle$ for several contagion timescales, with low (large)  values of $\mu$ 
corresponding to slow (fast) dynamics.
When contagion is fast compared to group reshuffling ($\mu \gg \omega$), the structure can be considered as effectively quenched: reinforcement persists within stable groups of adopters, and discontinuous transitions require large values of nonlinearities $\nu_c^\star$. Conversely, when the dynamics of group changes dominates ($\omega \gg \mu$), the system approaches the annealed regime where individuals experience many group configurations while they are in the adoption state, and $\nu_c^\star$ is reduced. 
\medskip

Finally, to compare the results found in different real-world systems, we rank them according to their effective structural coupling $Q$ (see definition in Eq.~(\ref{eq:Q_effective}) in Methods), a descriptor of the effective inter-group connectivity experienced during contagion. By construction, $Q$ captures that individual mobility across groups induces correlations between the dynamics of different groups, as the same individual can participate in multiple groups over its infectious period, therefore enhancing the effective inter-group connectivity. As shown in Fig.~\ref{fig:fig4}, weakly coupled systems require stronger collective reinforcement to undergo discontinuous transitions, whereas strongly coupled systems transition at lower reinforcement levels.
\medskip

Altogether, our results show that the type of phase transition observed is governed by the interplay between complex contagion and group dynamics and, in particular, by the time scales of the two processes. 
In this sense, the rate at which individuals switch groups directly modulates the effective coupling between groups and, consequently, the system-level behavior. As a result, static aggregated or annealed mean-field descriptions may fail to capture both the nature and the location of the critical boundary.

\medskip

\subsection{Discussion}

Temporality plays a fundamental role in shaping collective dynamics across a wide range of complex systems. While its impact has been explored in pairwise interaction networks, its consequences for group-based structures remain less understood.
Here, we introduced an analytically tractable model accounting for temporality through a group switching rate that captures how individuals change groups.
\medskip

We uncovered how temporality and collective reinforecement reorganize the entire phase diagram of higher-order contagion dynamics: they determine whether the phase transition from the absorbing to the active states is continuous, discontinuous, or absent, and induce a counterintuitive (non-monotonic) behavior of the invasion threshold giving rise to an optimal group switching rate. Introducing the group-size heterogeneity characteristic of real-world systems further enriches this picture. The interplay between temporality and heterogeneity yields multiple active states that differ in their degree of mesoscopic localization across groups. Therefore, new phases emerge, displaying hybrid transitions with continuous onset and active bistability, or discontinuous transitions with three coexisting stable states, both well beyond classical absorbing–active bistability. Importantly,
this behavior does not occur in static structures or mean-field approximations, demonstrating that temporality is not a correction to the structure but instead a primary dynamical mechanism. Although multistability has previously been reported in critical-mass dynamics, it was driven by the community structure of the static hypergraphs \cite{ferraz2023multistability}.
\medskip

Our results on real-world structures reveal that discontinuous transitions require substantially stronger collective reinforcement than suggested by static representations built from aggregated interaction data. Such aggregation suppresses temporal features \cite{iacopini2024temporal,gallo2024higher,arregui2024patterns} and can artificially alter hyperedge overlap and density \cite{lamata2025hyperedge,malizia2025hyperedge,kim2023contagion,malizia2025disentangling,burgio2024triadic,malizia2026nested}. As a consequence, both the location and qualitative nature of the transition may be mischaracterized when temporality is neglected.
\medskip

Another outcome of our work is the reconciliation of seemingly contradictory results. For linear contagion in homogeneous structures, temporality shifts the invasion threshold towards lower values by enhancing mixing \cite{valdano2015analytical,st2018phase}. This trend can reverse in higher-order contagions, but only under sufficiently strong nonlinear reinforcement or large group sizes. In these scenarios, rapid reshuffling prevents individuals from remaining long enough within stable groups to accumulate reinforcement, thereby hindering adoption. Supplementary Fig. 5 in Supplementary Note 8 shows that the consistent suppression of contagion reported by Chowdhary et al. \cite{chowdhary2021simplicial} arises when the reshuffling process does not preserve the degree sequence.
\medskip

Several avenues for future work emerge.  Although the role of temporality has been explored in consensus dynamics \cite{neuhauser2021consensus} and evolutionary game theory \cite{meng2025promoting,wang2026strategy}, its role in shaping phase transitions across other dynamical processes remains largely unexplored. Moreover, group switching itself may be adaptive: individuals modify their group participation in response to 
contagion \cite{gross2006epidemic,burgio2025characteristic,liu2025higher,mancastroppa2026higher} or in response to group traits \cite{stonge2024paradoxes, stonge2025defining}, a property that could be incorporated into solvable extensions of the present AME framework. 
More broadly, combining higher-order contagion with temporal heterogeneity in transmission, memory, or behavioural responses may yield richer dynamical transitions. 
Beyond these theoretical directions, our results raise methodological implications: aggregated hypergraphs can significantly misrepresent dynamical behaviour, and future empirical studies should integrate temporal information whenever possible.
\medskip

Together, these results establish a unified framework for temporal 
higher-order contagion, and reveal that temporality is not a secondary correction to higher-order contagion, but a primary mechanism that determines the macroscopic transition class. 
Ignoring temporality fundamentally mischaracterizes the nature and location of the critical transition, highlighting the need to treat time as an intrinsic part of the interaction structure.

\bibliography{biblio}

\section{Methods}

\subsection{The $\omega$AME model of complex contagion}

We follow the framework of St-Onge et al. \cite{st2022influential}, a non-linear generalization of the group-based Approximate Master Equations (AMEs) \cite{hebert2010propagation}, where we track the evolution of the susceptible population with membership $k$, denoted $s_k$, and the fraction of groups of size $n$ with $i$ infectious individuals within, denoted $f_{n,i}$. Adding the mechanism for group switching at rate $\omega$, the governing equations read:
\begin{align}
    \frac{ds_k}{dt}&=\mu(1-s_k)-krs_k\label{eq:sm}\\
    \frac{df_{n,i}}{dt}&=(i+1)\big(\mu + \omega (1-I)\big)\, f_{n,i+1}\nonumber\\&-\left[i(\mu+\omega (1-I))+(n-i)(\beta(n,i)+\rho+\omega I)\right]f_{n,i}\nonumber\\&+(n-i+1)(\beta(n,i-1)+\rho+\omega I)f_{n,i-1}. \label{eq:fni}
\end{align}

Eq. (\ref{eq:sm}) describes the change in the fraction of susceptible individuals belonging to $k$ groups simultaneously. This fraction increases when infected individuals recover and decreases when susceptibles become infected through exposure in any of the $k$ groups to which they belong.
\medskip

The terms in Eq. (\ref{eq:fni}) capture all transitions that modify the number of infectious individuals within a group of size $n$.  
The first term represents transitions from state \(i+1\) to \(i\), resulting either from the recovery of an infectious individual at rate $\mu$ or from its being swapped with a susceptible one at rate $\omega$.
The second term accounts for all events that remove probability mass from \(f_{n,i}\): any of the \(i\) infectious individuals may recover at rate $\mu$ or be swapped out at rate $\omega$, and any of the \(n-i\) susceptible individuals may become infected through within-group transmission at rate $\beta(n,i)$, external influence at rate $\rho$, or by being swapped with an infectious individual at rate $\omega$.  
The third term describes transitions from state \(i-1\) to \(i\), occurring when one of the \(n-(i-1)\) susceptibles becomes infectious via internal transmission at rate $\beta(n,i-1)$, external influence at rate $\rho$, or by being swapped with an infectious individual at rate $\omega$. Importantly, the external influence $\rho$ represents the mean-field infection pressure exerted by all groups to which a susceptible node belongs, excluding the focal group, and is given by
\begin{equation}
\rho(r) = r \cdot \frac{\sum_k k(k-1) s_k g_k}{\sum_k k s_k g_k},
\label{eq:rho}
\end{equation}
which corresponds to the product of the mean-field infection rate coming from a randomly selected external group and the mean excess membership of a susceptible node. In other words, if we pick a susceptible node in a given group, the prefactor represents the expected number of other groups to which it belongs. The mean-field infection rate itself is defined as 
\begin{equation}
r = \frac{\sum_{n,i} \beta(n,i)(n-i)f_{n,i}p_n}{\sum_{n,i} (n-i)f_{n,i}p_n},
\label{eq:r}
\end{equation}
which is the average value of $\beta(n,i)$ times the susceptible members within a group over the distribution of infected nodes. In addition, the global prevalence,
\begin{equation}
    I(t)=\sum_k(1-s_k(t)g_k),
    \label{eq:I(t)}
\end{equation}
serves as the main order parameter of the system.
Together, these contributions provide a compact characterization of the gain and loss processes that govern the evolution of \(f_{n,i}\).

\subsection{Stationary states}
In the stationary state, the whole system can be rewritten entirely in terms of the variables $r,\rho$ and $I$ (See Supplementary Note 1). We therefore define the functional 
\begin{equation}
    \mathcal{M}\left[\rho(r),I(r)\right]=\frac{\sum_{n,i} \beta(n,i)(n-i)f_{n,i}(\rho,I)p_n}{\sum_{n,i} (n-i)f_{n,i}(\rho,I)p_n},
    \label{eq:M}
\end{equation}
where the dependency on $\rho$ and $I$ enters through the stationary group-state distributions $f_{n,i}(\rho,I)$. At equilibrium, the system must satisfy the implicit self-consistency condition 
\begin{equation}
    r=\mathcal{M}\left[\rho(r),I(r)\right].
    \label{eq:stationary}
\end{equation}
This relation can be solved numerically to obtain the fixed points of the dynamics. To determine their stability, we additionally evaluate the Jacobian of the stationary system, as detailed in the Supplementary Note 1.

\subsection{Invasion threshold}

The invasion threshold corresponds to the critical points that mark the limit of the domain of validity of the solution $r=\mathcal{M}\left[\rho(r),I(r)\right]$, arising when $r$ is tangent to $\mathcal{M}\left[\rho(r),I(r)\right]$, i.e. when $\left.\frac{\partial \mathcal{M}}{\partial r}\right|_{r\rightarrow0}=1$. As derived in the Supplementary Note 2, the condition reads
\begin{eqnarray}
1&=&\left\langle \sum_{i=1}^{n}\;
\frac{n!}{(n-i-1)!\,i!}\; \left(\frac{\lambda_c}{\mu+\omega}\right)^{i}\; \prod_{j=1}^{\,i}j^{\nu}
\right\rangle\nonumber\\
&&\times\Bigg(\frac{\langle k(k-1)\rangle}{\langle k\rangle\langle n\rangle}+\frac{\omega}{\mu}\frac{\langle k\rangle}{\langle n\rangle}\Bigg).
\label{eq:inv_thres}
\end{eqnarray}

\subsection{Bistability threshold}

To fully characterize the phase portrait, we derive in the Supplementary Note 2 the condition for the tricritical point that separates the absorbing, active, and bistable regions. Imposing that $\left.\frac{\partial^2 \mathcal{M}}{\partial r^2}\right|_{r\rightarrow0}=0$ we reach that the condition reads
\begin{eqnarray}
0&=&F\frac{\langle k(k-1)\rangle^2}{\langle k\rangle^2}
+2G\frac{\langle k(k-1)\rangle}{\langle k\rangle}\frac{\langle k\rangle}{\mu}
+H\frac{\langle k\rangle^2}{\mu^2}\nonumber\\
&&+2\frac{\dfrac{1}{\mu}\Big(\langle k^2\rangle^2-\langle k^3\rangle \langle k\rangle\Big)-\dfrac{\omega}{\mu^2}\langle k^2\rangle\langle k\rangle^2}{\langle k(k-1)\rangle\langle k\rangle+\dfrac{\omega}{\mu}\langle k\rangle^3}.
\label{eq:bi_threshold}
\end{eqnarray}
Here, $F\equiv F\left[p_n,\beta(n,i),\omega\right]$, $G\equiv G\left[p_n,\beta(n,i),\omega\right]$, $H\equiv H\left[p_n,\beta(n,i),\omega\right]$ are specified in the Supplementary Note 2.

\subsection{Optimal group switching rate to maximize adoption}

The general expression for the critical line given $g(\kappa)=\delta_{\kappa,k}$ is derived in the Supplementary Note 3 and reads
\begin{equation}
\lambda_c(\omega;k)
= \frac{\mu+\omega}{2^{\nu}}
\left[
\sqrt{1 + \frac{2^{\nu}\mu}{k\omega+\mu(k-1)}} - 1
\right],
\label{eq:lambda_general_k}
\end{equation}
which reduces to Eq.~\eqref{eq:inv_thres_compact} for $k=1$. For $\nu>2$, differentiating Eq.~\eqref{eq:lambda_general_k} yields the position of the minimum:
\begin{equation}
\omega^\star(k)
= \mu\,
\frac{
2^{\nu/2+1} + 2^{\nu+1} - 4 - k(2^{\nu}-4)
}{
k(2^{\nu}-4)
},
\qquad (\nu>2).
\label{eq:wstar_general}
\end{equation}
As $k$ increases, the minimum $\omega^\star(k)$ shifts toward zero and eventually becomes non-positive. When this happens, the minimum disappears entirely and $\lambda_c(\omega;k)$ becomes strictly increasing in $\omega$.
\medskip

Evaluating Eq.~\eqref{eq:lambda_general_k} at $\omega^\star(k)$ gives the minimal threshold
\begin{equation}
\lambda_c^{\star}(k)
= \frac{1}{k}\,\lambda_c^{\star}(1)
= \frac{\mu}{k}\,\frac{2^{\nu/2}-1}{2^{\nu-1}}.
\label{eq:lamb_star_ge}
\end{equation}

\subsection{Persistence threshold}

The expression of the persistence threshold $\lambda_{p}(\omega)$ for large values of $\omega$, $g(k)=\delta_{k,1}$ and $p(n)=\delta_{n,3}$ is obtained in the Supplementary Note 5, and reads
\begin{equation}
\lambda_{p}(\omega)
=
\frac{
 -B_{p}(\omega)
 + \sqrt{
    B_{p}(\omega)^2
    - 4 A_{p}(\omega) C_{p}(\omega)
   }
}{
 2 A_{p}(\omega)
},
\label{eq:lambda-bi-final}
\end{equation}
where 
\begin{align}
A_{\mathrm{p}}(\omega,\mu)
&=
2^{\nu+1}\bigl(2^{\nu}\omega + 3\mu\bigr),
\\[1mm]
B_{\mathrm{p}}(\omega,\mu)
&=
4^{\nu}\omega^{2}
+ 2^{\nu+1}\mu\,\omega
+ 8\mu\,\omega
+ 9\mu^{2},
\\[1mm]
C_{\mathrm{p}}(\omega,\mu)
&=
-4\bigl(2^{\nu}-2\bigr)(\omega+\mu)^{2},
\label{eq:per_fin}
\end{align}
which match the stationary solutions of the AMEs for intermediate and large group switching rates, becoming exact in the annealed limit (see Supplementary Note 5). Its asymptotic value is
\begin{equation}
\lim_{\omega \to \infty} \lambda_{p,n=3}(\omega)
= \mu\frac{4\bigl(2^{\nu}-2\bigr)}{2^{2\nu}},
\label{eq:per_thres}
\end{equation}
and therefore, accounting for Eq. (\ref{eq:mean_field}), the width of the bistable region approaches 
\begin{equation}
\lim_{\omega \to \infty} \Delta\lambda_{n=3}(\omega)=
 \frac{\mu}{2}\left(1-2^{2-\nu}\right)^2.
\label{eq:bis_area}
\end{equation}

\subsection{Definition of the effective group switching rate}

To quantify the reshuffling process, we measure the average residence time of individuals within groups. Let $\tau_e$ denote the time interval between two consecutive group-change events involving the same individual (see Fig.~\ref{fig:Fig1}), and let $E$ be the total number of such events in the dataset. We define the effective group switching rate as the frequency of individual movement, computed as the inverse of the average inter-event time:
\begin{equation}
\langle \omega \rangle
= \frac{1}{\langle \tau \rangle},
\qquad
\langle \tau \rangle
= \frac{1}{E}\sum_{e=1}^{E} \tau_e, 
\label{eq:eff_plasticity}
\end{equation}
where an event is defined as the instantaneous switch of an individual from one group to another.
This quantity provides a direct measurement of how rapidly group composition changes in real systems. We refer the reader to the Supplementary Note 6 for further details.

\subsection*{Effective structural coupling}

Inspection of the invasion condition in Eq.~(\ref{eq:inv_thres}) shows that structural heterogeneity and temporal reshuffling enter the criticality criterion through a single structural–temporal combination. Therefore, we define an effective structural coupling $Q$, which integrates excess membership, group-size heterogeneity, and group switching:
\begin{equation}
Q
=
\left[
\frac{\langle k(k-1)\rangle}{\langle k\rangle}
+
\frac{\langle k\rangle\,\omega}{\mu}
\right]
\frac{\langle n(n-1)\rangle}{\langle n\rangle}.
\label{eq:Q_effective}
\end{equation}
The first term inside the brackets captures the structural excess membership of nodes, while the second term accounts for the enhancement induced by temporal reshuffling over the contagion timescale $\mu^{-1}$. The multiplicative factor involving $\langle n(n-1)\rangle$ encodes the excess group size, reflecting the number of potential reinforcement interactions within groups.

\section{Data availability}

The SocioPatterns datasets were downloaded from https://www.sociopatterns.org/datasets

\section{Code availability}

The code is available at https://github.com/santiagolaot/wAME

\section{Acknowledgements.} 
S.L.O. thanks M. Frasca for his help on finding the research question, and D. Soriano-Paños for insightful discussions on the interplay between dynamical processes at different characteristic timescales and his comments on the manuscript.  The authors acknowledge financial support from the Departamento de Industria e Innovaci\'on del Gobierno de Arag\'on y Fondo Social Europeo (FENOL group grant E36-23R, S.L.O and J.G.G.), from Ministerio de Ciencia e Innovaci\'on (grant PID2023-147734NB-I00, S.L.O and J.G.G.) from Gobierno de Aragón through a doctoral fellowship (S.L.O.), 
and from The National Science Foundation (award \#2419733, L.H.-D. and L.K.). F.M. acknowledges support from the Austrian Science Fund (FWF) through project 10.55776/PAT1652425.

\renewcommand{\figurename}{Supplementary Fig.}
\renewcommand{\tablename}{Supplementary Table}
\renewcommand{\theequation}{S.\arabic{equation}}

\setcounter{equation}{0}
\setcounter{figure}{0}

\onecolumngrid
\newpage

\section{Supplementary Note 1: Stable and unstable fixed points}

\subsection{Fixed point condition through detailed balance}

Imposing the stationary state condition in Eqs. (9)-(10) of the main text, i.e. that $ds_k/dt=0$ and $df_{n,i}/dt=0$, we obtain that
\begin{eqnarray}
    &&s_k=\frac{\mu}{\mu+mr},\\&&
    (i+1)\big(\mu + \omega(1-I)\big)\, f_{n,i+1}\nonumber\\&&=\left[i(\mu+\omega(1-I))+(n-i)(\beta(n,i)+\rho+\omega I)\right]f_{n,i}\nonumber\\&&-(n-i+1)(\beta(n,i-1)+\rho+\omega I)f_{n,i-1},
    \label{eq:fni}
\end{eqnarray}
where we assume the abuse of notation $s_k\equiv s_k^\star$ and $f_{n,i}\equiv f_{n,i}^\star$ for the shake of readability. We also know that in the stationary state $f_{n,i}$ must respect detailed balance:
\begin{equation}
(i+1)\big(\mu + \omega(1-I)\big)\, f_{n,i+1}
\;=\;
(n-i)\big(\beta(n,i) + \rho + \omega I\big)\, f_{n,i}.
\label{eq:DB}
\end{equation}
Incorporating this condition in Eq. (\ref{eq:fni}) leads to the iterative relation
\begin{equation}
    f_{n,i} \;=\;f_{n,0} \frac{n!}{(n-i)!\,i!}\;  \frac{\prod_{j=0}^{i-1}\left[\beta(n,j)+\rho+\omega I\right]}{(\mu+\omega(1-I))^{i}},
    \label{eq:recursive}
\end{equation}

where $f_{n,0}=1-\sum_{i=1}^n f_{n,i}$ due to the normalization constraint.
\medskip

In the stationary state, the whole system can be rewritten in terms of $r,\rho$ and $I$. Therefore, we can define a function $\mathcal{M}\left[\rho(r),I(r)\right]$, reading
\begin{equation}
    \mathcal{M}\left[\rho(r),I(r)\right]=\frac{\sum_{n,i} \beta(n,i)(n-i)f_{n,i}(\rho,I)p_n}{\sum_{n,i} (n-i)f_{n,i}(\rho,I)p_n},
    \label{eq:M}
\end{equation}
where the dependency on $\rho$ and $I$ is in the $f_{n,i}(\rho,I)$, and
that must fulfill the implicit relation $r=\mathcal{M}\left[\rho(r),I(r)\right]$. This relation can be used to solve numerically for the fixed points.

\subsection{Jacobian of the dynamics}

To evaluate the stability of the fixed points, we must also look at their Jacobian
\begin{equation}
    J_r=\frac{d\mathcal{M}}{dr}=\frac{\partial \mathcal{M}}{\partial\rho}\frac{\partial \rho}{\partial r}+\frac{\partial \mathcal{M}}{\partial I}\frac{\partial I}{\partial r}.
\end{equation}
In the former expression, the derivatives $\frac{\partial \rho}{\partial r}$ and $\frac{\partial I}{\partial r}$
can be straightforwardly computed as
\begin{eqnarray}
    \frac{\partial \rho}{\partial r}&=&\frac{u_2}{v_2}+r\left[\frac{1}{v_2}\frac{du_2}{dr}-\frac{u_2}{v_2^2}\frac{dv_2}{dr}\right],\\
    \frac{\partial I}{\partial r}&=&-\sum_kk\frac{ds_k}{dr}g_k,
\end{eqnarray}
where
\begin{eqnarray}
    u_2&=&\sum_k k(k-1) s_k(r) g_k,\nonumber\\
    v_2&=&\sum_k k s_k(r) g_k,\nonumber\\
    \frac{du_2}{dr}&=&\sum_k k(k-1) \frac{ds_k}{dr} g_k,\nonumber\\
    \frac{dv_2}{dr}&=&\sum_k k \frac{ds_k}{dr} g_k,\nonumber\\
    \frac{ds_k}{dr}&=&-\frac{\mu k}{(\mu+kr)^2}.\nonumber
\end{eqnarray}
However, the derivatives $\frac{\partial \mathcal{M}}{\partial\rho}$ and $\frac{\partial \mathcal{M}}{\partial I}$ are more difficult to be computed as they explicitly depend on $\frac{df_{n,i}}{d\rho}$ and $\frac{df_{n,i}}{dI}$ as
\begin{eqnarray}
    \frac{\partial \mathcal{M}}{\partial\rho}&=&\frac{1}{v_1}\frac{du_1}{d\rho}-\frac{u_1}{v_1^2}\frac{dv_1}{d\rho},\\
    \frac{\partial \mathcal{M}}{\partial I}&=&\frac{1}{v_1}\frac{du_1}{dI}-\frac{u_1}{v_1^2}\frac{dv_1}{dI},\\
\end{eqnarray}
where
\begin{eqnarray}
    u_1&=&\sum_{n,i} \beta(n,i)(n-i)f_{n,i}(\rho,I)p_n,\nonumber\\
    v_1&=&\sum_{n,i} (n-i)f_{n,i}(\rho,I)p_n,\nonumber\\
    \frac{du_1}{d\rho}&=&\sum_{n,i}\beta(n,i)(n-i)\frac{df_{n,i}}{d\rho}p_n,\nonumber\\
\frac{du_1}{dI}&=&\sum_{n,i}\beta(n,i)(n-i)\frac{df_{n,i}}{dI}p_n,\nonumber\\
\frac{dv_1}{d\rho}&=&\sum_{n,i} (n-i)\frac{df_{n,i}}{d\rho}p_n,\nonumber\\
\frac{dv_1}{dI}&=&\sum_{n,i} (n-i)\frac{df_{n,i}}{dI}p_n\nonumber.
\end{eqnarray}
In order to compute the derivatives $\frac{df_{n,i}}{d\rho}$ and $\frac{df_{n,i}}{dI}$, we rewrite the stationary recursive relation in terms of a closed form and its recursive factor. From Eq. (\ref{eq:recursive}) we first define $B_j=\beta(n,j)+\rho+\omega I$ and $A=\mu+\omega(1-I)$, yielding 
\begin{eqnarray}
    f_{n,i}=f_{n,0}\binom{n}{i}\prod_{j=0}^{i-1}\frac{B_j}{A}\;\;\Leftrightarrow \;\;f_{n,i}=\frac{g_{n,i}}{Z_n},
\end{eqnarray}
with 
\begin{eqnarray}
    g_{n,i}&=&\binom{n}{i}\prod_{j=0}^{i-1}\frac{B_j}{A},\;\;\;\;\;\;Z=\sum_{l=0}^kg_{n,l},\label{eq:gni}
\end{eqnarray}
where we have set as normalization $g_{n,0}=1$ (since $f_{n,0}$ is defined by normalization). Therefore, we can express the desired derivatives in terms of $\theta=\{\rho,I\}$ as
\begin{equation}
\frac{df_{n,i}}{d\theta}=\frac{g_{n,i}'Z_n-g_{n,i}Z_n'}{Z_n^2}=f_{n,i}\left(\frac{g_{n,i}'}{g_{n,i}}-\frac{Z_n'}{Z_n}\right)=f_{n,i}\left(\frac{g_{n,i}'}{g_{n,i}}-\sum_l f_{n,l}\frac{g_{n,l}'}{g_{n,l}}\right)=f_{n,i}\left(\Delta_{n,i}^{\theta}-\sum_l f_{n,l}\Delta_{n,l}^{\theta}\right),
\end{equation}
where we have used that $\frac{Z_n'}{Z_n}=\frac{\sum_lg_{n,l}'}{\sum_lg_{n,l}}=\sum_l\frac{g_{n,l}}{Z_n}\frac{g_{n,l}'}{g_{n,l}}=\sum_lf_{n,l}\frac{g_{n,l}'}{g_{n,l}}$, and we have defined
\begin{equation}
    \Delta_{n,i}^{\theta}\equiv\frac{g'}{g}=\frac{1}{g}\frac{dg}{d\theta}=\frac{d}{d\theta}\text{ln}g_{n,i}.
\end{equation}
Expressing the former quantity in terms of the derivative of the logarithm eases the mathematical derivation, as it allows for splitting the terms of Eq. (\ref{eq:gni}) as follows:
\begin{eqnarray}
    \text{ln}(g_{n,i})&=&\text{ln}\binom{n}{i}+\sum_{j=0}^{i-1}\text{ln}B_j-i\text{ln}A,\\
    \frac{d}{d\theta}\text{ln}(g_{n,i})&=&\sum_{j=0}^{i-1}\frac{1}{B_j}\frac{dB_j}{d\theta}-i\frac{1}{A}\frac{dA}{d\theta},
\end{eqnarray}
being $\frac{dB_j}{d\rho}=1$, $\frac{dB_j}{dI}=\omega$, $\frac{dA}{d\rho}=0$ and $\frac{dA}{dI}=-\omega$. Therefore, we reach both closed expressions for $\frac{df_{n,i}}{d\theta}$, $\theta=\{\rho,I\}$ as
\begin{eqnarray}
    \frac{df_{n,i}}{d\theta}&=&f_{n,i}\left(\Delta_{n,i}^{\theta}-\sum_l f_{n,l}\Delta_{n,l}^{\theta}\right)\;\;\;\text{with}\;\;\;\theta=\{\rho,I\},\\
    \Delta_{n,i}^{\rho}&=&\sum_{j=0}^{i-1}\frac{1}{\beta(n,j)+\rho+\omega I},\\
\Delta_{n,i}^{I}&=&\omega\Delta_{n,i}^{\rho}+\frac{\omega i}{\mu+\omega(1-I)}.
\end{eqnarray}
\medskip

The expressions above provide all the required ingredients to evaluate
the Jacobian $J_r$ at any stationary solution of the AME system. The sign of $J_r$ fully determines the
linear stability of the corresponding fixed point: $J_r<0$ identifies a
stable branch, $J_r>0$ an unstable one, and the condition $J_r=0$
detects the onset of saddle–node bifurcations.

\newpage

\section{Supplementary Note 2: Conditions for the invasion threshold and the tricritical point}

The invasion threshold is the critical point at which the non-trivial solution $r=\mathcal{M}\left[\rho(r),I(r)\right]$ first emerges. This occurs when $\mathcal{M}\left[\rho(r),I(r)\right]$ is tangent to $r$, i.e. when
\begin{eqnarray}
    \frac{d\mathcal{M}}{dr}=\frac{\partial \mathcal{M}}{\partial\rho}\frac{\partial \rho}{\partial r}+\frac{\partial \mathcal{M}}{\partial I}\frac{\partial I}{\partial r}=1.
    \label{eq:invasion_cond}
\end{eqnarray}
Moreover, to have tricritical points (i.e., where absorbing, active, and bistable phases merge), the nontrivial solutions must be degenerated, which requires that the second derivative is null.
\begin{equation}
\frac{d^2\mathcal{M}}{dr^2}
=
\frac{\partial^2 \mathcal{M}}{\partial\rho^2}\Big(\frac{d\rho}{dr}\Big)^2
+2\frac{\partial^2 \mathcal{M}}{\partial\rho\partial I}\frac{d\rho}{dr}\frac{dI}{dr}
+\frac{\partial^2 \mathcal{M}}{\partial I^2}\Big(\frac{dI}{dr}\Big)^2
+\frac{\partial \mathcal{M}}{\partial\rho}\frac{d^2\rho}{dr^2}
+\frac{\partial \mathcal{M}}{\partial I}\frac{d^2I}{dr^2}=0.
\label{eq:tri_cond}
\end{equation}

\subsection{Derivatives}

In the limit of $r\rightarrow0$, it is possible to derive a semi-analytical expression for both the invasion threshold and the tricritical point. In this limit, $s_k\rightarrow1$, $\rho\rightarrow0$, $I\rightarrow0$ and $f_{n,i}\rightarrow\delta_{i,0}$, i.e. all nodes are susceptible. 
Some of the derivatives are simple, as
\begin{eqnarray}
\left.\frac{d\rho}{dr}\right|_{(0,0)} &=& \frac{\langle k(k-1)\rangle}{\langle k\rangle},\\
\left.\frac{dI}{dr}\right|_{(0,0)}&=& \frac{\langle k\rangle}{\mu},\\
\left.\frac{d^2\rho}{dr^2}\right|_{(0,0)} &=& \frac{2}{\mu}\left(\frac{\langle k^2\rangle^2}{\langle k\rangle^2}-\frac{\langle k^3\rangle}{\langle k\rangle}\right),\\
\left.\frac{d^2I}{dr^2}\right|_{(0,0)}&=& -2\frac{\langle k^2\rangle}{\mu^2},\\
\end{eqnarray}
\medskip

However, to solve the core derivatives $\frac{\partial \mathcal{M}}{\partial\rho}|_{r\rightarrow0}$ and $\frac{\partial \mathcal{M}}{\partial I}|_{r\rightarrow0}$ we realize that the denominator of 
\begin{equation}
    \mathcal{M}\left[\rho(r),I(r)\right]=\frac{\sum_{n,i} \beta(n,i)(n-i)f_{n,i}(\rho,I)p_n}{\sum_{n,i} (n-i)f_{n,i}(\rho,I)p_n},
\end{equation}
becomes $\langle n\rangle$, and in the numerator all terms are small, since $\beta(n,0)=0$, which is the one that corresponds to $\delta_{i,0}$. Therefore, we approximate that
\begin{equation}
    \frac{d\mathcal{M}}{d\theta}|_{r\rightarrow0}=\frac{1}{\langle n \rangle}\sum_{n,i} \beta(n,i)(n-i)p_n h_{n,i}^{\theta},\;\;\;\;\; \text{with}\;\;h_{n,i}^{\theta}=\frac{f_{n,i}}{d\theta}|_{r\rightarrow0},\;\;\;\theta=\{\rho,I\}.
\end{equation}
For the second derivative we have that 
\begin{eqnarray}
\frac{d^2\mathcal{M}}{d\theta^2} 
&=& \frac{(u'' v + u' v' - u' v' - u v'') v^2 
- (u' v - u v') 2 v v'}{v^4}\\
&=& \frac{v \left(u'' v - u v'' \right) 
+ 2 v' \left(u v' - u' v \right)}{v^3}.
\end{eqnarray}
Now, in the limit $\rho \to 0$ we have $u(0) = 0$ and $v(0) = \langle n \rangle$. Substituting these values, all terms proportional to $u$ vanish, leaving
\begin{eqnarray}
   \left. \frac{d^2 \mathcal{M}}{d\theta^2} \right|_{(0,0)}
&=& \frac{u'' v^2 - 2 v u' v'}{v^3}\\
&=&\frac{1}{\langle n \rangle} \frac{d^2 u}{d\theta^2}
- \frac{2}{\langle n \rangle^2} 
\frac{du}{d\theta}
\frac{dv}{d\theta},
\end{eqnarray}
where, after defining $h^\theta_{n,i}=\frac{df_{n,i}}{d\theta}$ and $l^\theta_{n,i}=\frac{d^2f_{n,i}}{d\theta^2}$:
\begin{eqnarray}
    \frac{du}{d\theta}&=&\sum_{n,i}\beta(n,i)(n-i)p_nh^\theta_{n,i},\nonumber\\
    \frac{dv}{d\theta}&=&\sum_{n,i} (n-i)p_nh^\theta_{n,i},\nonumber\\
    \frac{d^2u}{d\theta^2}&=&\sum_{n,i}\beta(n,i)(n-i)p_nl^\theta_{n,i}.\nonumber\\
\end{eqnarray}

Moreover, the derivative $ \left. \frac{d^2 \mathcal{M}}{d\rho dI} \right|_{(0,0)}$ requires special treatment, as
\begin{eqnarray}
\left. \frac{d^2 \mathcal{M}}{d\rho dI} \right|_{(0,0)}
&=&\frac{1}{\langle n \rangle} \frac{d^2 u}{d\rho d\theta}
- \frac{1}{\langle n \rangle^2} \left(
\frac{du}{d\rho}
\frac{dv}{dI}+\frac{du}{d I}
\frac{dv}{d\rho}\right),
\end{eqnarray}
and therefore we define $l^{\text{mix}}_{n,i}=\frac{d^2f_{n,i}}{d\rho dI}$ and
\begin{eqnarray}
    \frac{d^2u}{d\rho dI}&=&\sum_{n,i}\beta(n,i)(n-i)p_nl^{\text{mix}}_{n,i}.\nonumber\\
\end{eqnarray}
\medskip

Overall, we need to compute $h^\rho_{n,i},h^I_{n,i},l^\rho_{n,i},l^I_{n,i},l^{\text{mix}}_{n,i}$. In order to do so we start from the stationary relation between the $f_{n,i}$'s reported in Eq. (\ref{eq:fni}), knowing also that in the stationary state $f_{n,i}$ must respect detailed balance condition in Eq. (\ref{eq:DB}). Moreover, since probabilities must satisfy $\sum_i f_{n,i}=1$ for each $n$, the derivatives fulfill
\begin{equation}
\sum_i h_{n,i}^\theta = 0, 
\qquad 
\sum_i l_{n,i}^\theta = 0,
\end{equation}
so that the base values can be written as
\begin{equation}
h_{n,0}^\theta = -\sum_{i\ge1} h_{n,i}^\theta,
\qquad
l_{n,0}^\theta = -\sum_{i\ge1} l_{n,i}^\theta,
\qquad
\theta\in\{\rho,I,\text{mix}\}.
\end{equation}

\noindent\textit{First derivative in terms of $\rho$: $h^\rho_{n,i}$}. The derivative reads
\begin{eqnarray}
    &&
    (i+1)\big(\mu + \omega(1-I)\big)\, h_{n,i+1}^{\rho}\nonumber\\&&=(n-i)f_{n,i}+\left[i(\mu+w(1-I))+(n-i)(\beta(n,i)+\rho+\omega I)\right]h_{n,i}^{\rho}\nonumber\\&&-(n-i+1)f_{n,i-1}-(n-i+1)(\beta(n,i-1)+\rho+\omega I)h_{n,i-1}^{\rho},
\end{eqnarray}
and after the detailed balance and substituting that $\rho=I=f_{n}=0$ we obtain that:
\begin{itemize}
    \item if $i=0$: $h_{n,1}^{\rho}=\frac{n}{\mu+\omega}$
    \item if $i>0$: $h_{n,i+1}^{\rho}=\frac{(n-i)}{(i+1)}\frac{\beta(n,i)}{(\mu+\omega)}h_{n,i}^{\rho}$
\end{itemize}
\medskip

\noindent\textit{First derivative in terms of $I$: $h^I_{n,i}$}. The derivative reads
\begin{eqnarray}
    &&
    (i+1)(-\omega)f_{n,i+1}+(i+1)\big(\mu + \omega(1-I)\big)\, h_{n,i+1}^{I}\nonumber\\&&=(-i\omega+(n-i)\omega)f_{n,i}+\left[i(\mu+\omega(1-I))+(n-i)(\beta(n,i)+\rho+\omega I)\right]h_{n,i}^{I}\nonumber\\&&-(n-i+1)\omega f_{n,i-1}-(n-i+1)(\beta(n,i-1)+\rho+\omega I)h_{n,i-1}^{I},
\end{eqnarray}
and after using detailed balance and substituting that $\rho=I=f_{n}=0$ we obtain that:
\begin{itemize}
    \item if $i=0$: $h_{n,1}^{I}=\frac{n\omega}{\mu+\omega}$
    \item if $i>0$: $h_{n,i+1}^{I}=\frac{(n-i)}{(i+1)}\frac{\beta(n,i)}{(\mu+\omega)}h_{n,i}^{I}$
\end{itemize}
\medskip

\noindent\textit{Second derivative in terms of $\rho$: $l^\rho_{n,i}$}. The derivative reads
\begin{eqnarray}
    &&
    (i+1)\big(\mu + \omega(1-I)\big)\, l_{n,i+1}^{\rho}\nonumber\\&&=2(n-i)h_{n,i}^{\rho}-2(n-i+1)h_{n,i-1}^{\rho}+\left[i(\mu+\omega(1-I))+(n-i)(\beta(n,i)+\rho+\omega I)\right]l_{n,i}^{\rho}\nonumber\\&&-(n-i+1)(\beta(n,i-1)+\rho+\omega I)l_{n,i-1}^{\rho},
\end{eqnarray}
and substituting that $\rho=I=f_{n}=0$ we obtain that:
\begin{itemize}
    \item if $i=0$: $l_{n,1}^{\rho}=\frac{2n}{\mu+\omega}h_{n,0}^{\rho}$
    \item if $i>0$: 
    \begin{eqnarray}
        (i+1)\big(\mu + \omega\big)\, l_{n,i+1}^{\rho}&=&2(n-i)h_{n,i}^{\rho}-2(n-i+1)h_{n,i-1}^{\rho}\nonumber\\&&+\left[i(\mu+\omega)+(n-i)\beta(n,i)\right]l_{n,i}^{\rho}\nonumber\\&&-(n-i+1)\beta(n,i-1)l_{n,i-1}^{\rho},\nonumber
    \end{eqnarray}
\end{itemize}
\medskip

\noindent\textit{Second derivative in terms of $I$: $l^I_{n,i}$}. The derivative reads
\begin{eqnarray}
    &&
    -2\omega(i+1)h_{n,i+1}^I+(i+1)\big(\mu + \omega(1-I)\big)\, l_{n,i+1}^{I}\nonumber\\&&=2(-i\omega+(n-i)\omega)h_{n,i}^{I}-2(n-i+1)h_{n,i-1}^{I}+\left[i(\mu+\omega(1-I))+(n-i)(\beta(n,i)+\rho+\omega I)\right]l_{n,i}^{I}\nonumber\\&&-2(n-i+1)\omega h_{n,i-1}^I-(n-i+1)(\beta(n,i-1)+\rho+\omega I)l_{n,i-1}^{I},
\end{eqnarray}
and substituting that $\rho=I=f_{n}=0$ we obtain that:
\begin{itemize}
    \item if $i=0$: $l_{n,1}^{I}=\frac{2\omega}{\mu+\omega}\left(nh_{n,0}^{I}+h_{n,1}^I\right)$
    \item if $i>0$: 
    \begin{eqnarray}
        (i+1)\big(\mu + \omega\big)\, l_{n,i+1}^{I}&=&2\omega\left[-i+(n-i)\right]h_{n,i}^{I}-2\omega(n-i+1)h_{n,i-1}^{I}\nonumber\\&&+\left[i(\mu+\omega)+(n-i)\beta(n,i)\right]l_{n,i}^{I}\nonumber\\&&-(n-i+1)\beta(n,i-1)l_{n,i-1}^{I}\nonumber\\
        &&+2\omega(i+1)h_{n,i+1}^{I},\nonumber
    \end{eqnarray}
\end{itemize}
\medskip

\noindent\textit{Second derivative in terms of $I$ and $\rho$ (mixed derivative): $l^{\text{mix}}_{n,i}$}. The derivative reads
\begin{eqnarray}
    &&
    -\omega(i+1)h_{n,i+1}^\rho+(i+1)\big(\mu + \omega(1-I)\big)\, l_{n,i+1}^{\text{mix}}\nonumber\\&&=(n-i)h_{n,i}^{I}+\omega\left[-i+(n-i)\right]h_{n,i}^{\rho}+\left[i(\mu+\omega(1-I))+(n-i)(\beta(n,i)+\rho+\omega I)\right]l_{n,i}^{\text{mix}}\nonumber\\&&-(n-i+1)\left[h_{n,i-1}^{I}+\omega h_{n,i-1}^{\rho}\right]-(n-i+1)(\beta(n,i-1)+\rho+\omega I)l_{n,i-1}^{\text{mix}},
\end{eqnarray}
and substituting that $\rho=I=f_{n}=0$ we obtain that:
\begin{itemize}
    \item if $i=0$: $l_{n,1}^{\text{mix}}=\frac{1}{\mu+\omega}\left(nh_{n,0}^{I}+n\omega h_{n,0}^{\rho}+\omega h_{n,1}^\rho\right)$
    \item if $i>0$: 
    \begin{eqnarray}
        (i+1)\big(\mu + \omega\big)\, l_{n,i+1}^{\text{mix}}&=&(n-i)h_{n,i}^I+\omega\left[-i+(n-i)\right]h_{n,i}^{\rho}-(n-i+1)\left[h_{n,i-1}^{I}+wh_{n,i-1}^{\rho}\right]\nonumber\\&&+\left[i(\mu+\omega)+(n-i)\beta(n,i)\right]l_{n,i}^{\text{mix}}\nonumber\\&&-(n-i+1)\beta(n,i-1)l_{n,i-1}^{\text{mix}}\nonumber\\
        &&+\omega(i+1)h_{n,i+1}^{\rho},\nonumber
    \end{eqnarray}
\end{itemize}

\subsection{General expressions}

Incorporating the expressions of the derivatives into Eq. (\ref{eq:invasion_cond}), we obtain the following closed expression for the invasion threshold:
\begin{equation}
\left\langle \sum_{i=1}^{n}\;
\frac{n!}{(n-i-1)!\,i!}\; (\mu+\omega)^{-i}\; \prod_{j=1}^{\,i}\beta(n,j)
\right\rangle
\Bigg(\frac{\langle k(k-1)\rangle}{\langle k\rangle\langle n\rangle}+\frac{\omega}{\mu}\frac{\langle k\rangle}{\langle n\rangle}\Bigg)
=1.%
\label{eq:invasion_threshold}
\end{equation}
\medskip

Similarly, incorporating the expressions of the derivatives into Eq. (\ref{eq:tri_cond}), we obtain the following closed expression for the tricritical point.
\begin{equation}
F\frac{\langle k(k-1)\rangle^2}{\langle k\rangle^2}
+2G\frac{\langle k(k-1)\rangle}{\langle k\rangle}\frac{\langle k\rangle}{\mu}
+H\frac{\langle k\rangle^2}{\mu^2}
+J\left[\frac{2}{\mu}\left(\frac{\langle k^2\rangle^2}{\langle k\rangle^2}-\frac{\langle k^3\rangle}{\langle k\rangle}\right)-2\omega\frac{\langle k^2\rangle}{\mu^2}\right]=0.
\end{equation}

where $F\equiv F\left[p_n,\beta(n,i),\omega\right]=\left.\frac{\partial^2 M}{\partial\rho^2}\right|_{(\rho,I) \to (0,0)}$, $G\equiv G\left[p_n,\beta(n,i),\omega\right]=\left.\frac{\partial^2 M}{\partial\rho\partial I}\right|_{(\rho,I) \to (0,0)}$, $H\equiv H\left[p_n,\beta(n,i),\omega\right]=\left.\frac{\partial^2 M}{\partial I^2}\right|_{(\rho,I) \to (0,0)}$, $J\equiv J\left[p_n,\beta(n,i),\omega\right]=\left.\frac{\partial M}{\partial\rho}\right|_{(\rho,I) \to (0,0)}$. Recalling from the invasion threshold that $\left.\frac{\partial M}{\partial\rho}\right|_{(\rho,I) \to (0,0)}=\left(\frac{\partial \rho}{\partial r}+\omega\frac{\partial I}{\partial r}\right)^{-1}$, we can rewrite the expression as
\begin{eqnarray}
0&=&F\frac{\langle k(k-1)\rangle^2}{\langle k\rangle^2}
+2G\frac{\langle k(k-1)\rangle}{\langle k\rangle}\frac{\langle k\rangle}{\mu}
+H\frac{\langle k\rangle^2}{\mu^2}+2\frac{\dfrac{1}{\mu}\Big(\langle k^2\rangle^2-\langle k^3\rangle \langle k\rangle\Big)-\dfrac{\omega}{\mu^2}\langle k^2\rangle\langle k\rangle^2}{\langle k(k-1)\rangle\langle k\rangle+\dfrac{\omega}{\mu}\langle k\rangle^3}.
\end{eqnarray}

\newpage

\section{Supplementary Note 3: Invasion threshold}

\subsection{Limiting cases of the invasion threshold}

We can obtain an explicit expression of the invasion threshold by imposing that $i=1$ in Eq. (\ref{eq:invasion_threshold}), i.e. assuming that at the onset of the epidemic all infectious groups have only one infectious individual. The obtained expression reads
\begin{equation}
\lambda_c \approx 
\frac{\mu + \omega}{\langle n(n-1)\rangle}
\left[
\dfrac{\langle k(k-1)\rangle}{\langle k\rangle \langle n\rangle} 
+ \dfrac{\omega}{\mu}\dfrac{\langle k\rangle}{\langle n\rangle}
\right]^{-1}.
\label{eq:lambdac_supp}
\end{equation}

\subsubsection{Quenched limit}

We now derive the quenched limit reported in Eq. (1) of the main text for the case in which each individual participates in a single group at a time ($g(k)=\delta_{k,1}$). In this case $\langle k\rangle = 1$ and $\langle k(k-1)\rangle = 0$.
Substituting these relations into Eq.~\eqref{eq:lambdac_supp}, the expression
inside the brackets simplifies to $\omega/(\mu\,\langle n\rangle)$.
Equation~\eqref{eq:lambdac_supp} thus becomes
\begin{equation}
\lambda_c \approx 
\frac{\mu + \omega}{\langle n(n-1)\rangle}
\cdot
\frac{\mu\,\langle n\rangle}{\omega}.
\end{equation}
In the limit $\omega \to 0$, the numerator satisfies $\mu + \omega \sim \mu$, so that
\begin{equation}
\lambda_c \sim
\frac{\mu^2\,\langle n\rangle}{\omega\,\langle n(n-1)\rangle},
\qquad
\text{as } \omega \to 0.
\end{equation}
Therefore, the invasion threshold diverges in the quenched limit,
\begin{equation}
\lim_{\omega \to 0} \lambda_c(\omega)= \infty,
\end{equation}
showing that infinitesimal adoption cannot invade when the higher-order structure is static and fragmented (each individual belongs to a single group only).

\subsubsection{Annealed limit}

To compute the annealed (mean-field) limit, we identify the dominant terms in Eq. (\ref{eq:lambdac_supp}) as $\omega \to \infty$.  
In the numerator, $\mu + \omega \sim \omega$, while the term inside the brackets becomes $\omega\langle k\rangle/(\mu\langle n\rangle)$. Substituting these asymptotic forms into Eq.~\eqref{eq:lambdac_supp} yields to the invasion threshold in the annealed limit
\begin{equation}
\lim_{\omega \to \infty} \lambda_c(\omega)
=
\frac{\mu\, \langle n\rangle}{\langle k\rangle\, \langle n(n-1)\rangle},
\label{eq:lc_lim}
\end{equation}
that corresponds to the mean-field limit of the dynamics.

\subsection{Closed expression of the invasion threshold}

We start from the invasion threshold condition in Eq. (\ref{eq:invasion_threshold}), and consider the case of fixed group size ($p(n')=\delta_{n',n}$), so that 
$\langle n\rangle = n$ and the angular brackets over $n$ become trivial. Defining $x \equiv \lambda_c/(\mu+\omega)$, and using $
\prod_{j=1}^i j^\nu = (i!)^\nu$ and $\frac{n!}{(n-i-1)!\,i!}
= n\,\binom{n-1}{i}$ for $i=1,...,n-1$, we can define the polynomial
\begin{equation}
F_n(x)
=
n \sum_{i=1}^{n-1}
\binom{n-1}{i}\,(i!)^\nu\,x^i,
\label{eq:Fn-def-k}
\end{equation}
\medskip
so that Eq.~\eqref{eq:invasion_threshold}
becomes
\begin{equation}
F_n(x)
=
\left[\dfrac{\langle k(k-1)\rangle}{\langle k\rangle\,n}
+
\dfrac{\omega}{\mu}\dfrac{\langle k\rangle}{n}
\right]^{-1},
\label{eq:Fn-eq-k}
\end{equation}
which is a $(n-1)$-degree
polynomial equation for $x$ which implicity defines the invasion threshold via $\lambda_c=x(\mu+\omega)$. Note that starting from Eq. (\ref{eq:Fn-eq-k}) we can recover the quenched and annealed limits.
\medskip

For fixed group size $p(n)=\delta_{n,3}$ and fixed membership $g(\kappa)=\delta_{\kappa,k}$,
the threshold condition in Eq. (\ref{eq:Fn-eq-k}) reduces to 
\begin{equation}
x\bigl(2^{\nu}x+2\bigr)
= \frac{\mu}{k\omega+\mu(k-1)}.
\end{equation}
Solving this quadratic equation for \(x\) keeping the physical
(positive) root and recovering that  $\lambda_c=x(\mu+\omega)$,  we obtain the critical line
\begin{equation}
\lambda_c(\omega;k)
= \frac{\mu+\omega}{2^{\nu}}
\left[
\sqrt{1 + \frac{2^{\nu}\mu}{k\omega+\mu(k-1)}} - 1
\right].
\label{eq:lambdac_general_k}
\end{equation}

\subsection*{Optimal group switching for maximizing adoption}

To locate the minimum of the critical line we differentiate
Eq.~\eqref{eq:lambdac_general_k} with respect to \(\omega\)
and set the derivative to zero. To do so, we can define $D(\omega)=(k-1)+\frac{k\omega}{\mu}$ and $Y(\omega)=\sqrt{1+\frac{2^{\nu}}{D(\omega)}}$.
Differentiating Eq.~\eqref{eq:lambdac_general_k} gives
\[
\frac{d\lambda_c}{d\omega}
=
\frac{1}{2^{\nu}}
\left[
Y-1
-
\frac{(\mu+\omega)k}{2\mu\,D(\omega)^2\,Y}
\right].
\]
The minimum of $\lambda_c(\omega;k)$ satisfies 
$d\lambda_c/d\omega=0$, which leads to $2\mu D(\omega)\,Y(\omega)=k(\mu+\omega)\bigl[Y(\omega)+1\bigr]$. Finally, using the identity $Y^2=1+\frac{2^{\nu}}{D}$, and substituting $D(\omega)=(k-1)+k\omega/\mu$, we obtain an algebraic
equation whose unique positive solution yields the optimal mixing rate,
\begin{equation}
\omega^{\star}(k)
= \mu\,
\frac{2^{\nu/2+1} + 2^{\nu+1} - 4 - k\,(2^{\nu}-4)}
     {k\,(2^{\nu}-4)},
\qquad (\nu>2),
\label{eq:omega_opt}
\end{equation}
Evaluating Eq.~\eqref{eq:lambdac_general_k} at \(\omega^{\star}(k)\)
gives the minimal value of the critical line,
\begin{equation}
\lambda_c^{\star}(k)
= \frac{1}{k}\,\lambda_c^{\star}(1)
= \frac{\mu}{k}\,\frac{2^{\nu/2}-1}{2^{\nu-1}}.
\label{eq:lambda_min}
\end{equation}

\newpage

\section{Supplementary Note 4: Effect of membership and group size on the critical behaviour}

\subsection{Effect of membership}

\begin{figure*}[h]
    \centering
    \includegraphics[width=0.98\linewidth]{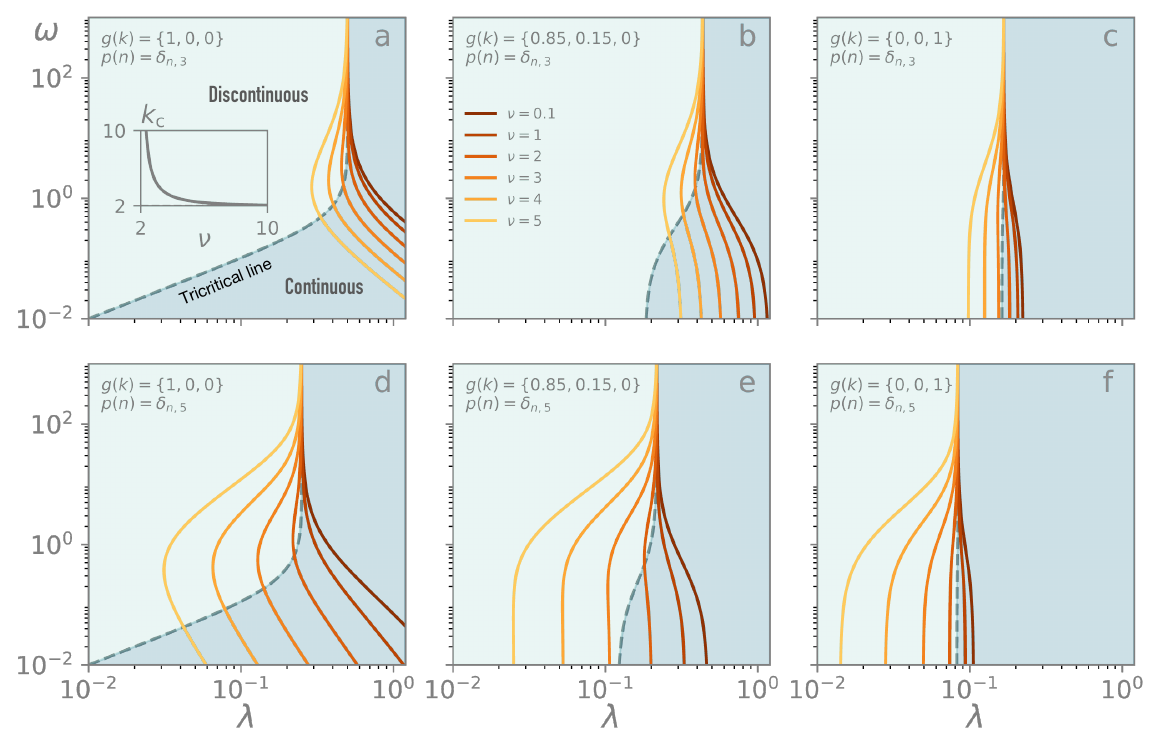}
   \caption{\textbf{Interplay between membership, group size, group switching, and synergy reshapes the phase diagram.}
    Phase portraits showing the critical adoption threshold $\lambda_c$ as a function of the group switching rate $\omega$ for different synergy factors, group sizes, and membership distributions. Panels \textbf{a}–\textbf{c} (\textbf{d}–\textbf{f}) correspond to fixed group size $p(n)=\delta_{n,3}$ ($p(n)=\delta_{n,5}$), while the left, central, and right columns correspond to $\langle k \rangle = 1.00$, $1.15$, and $3.00$, respectively. 
    Each panel highlights the tricritical lines separating regions with continuous and discontinuous transitions. 
    The inset in panel \textbf{a} shows the critical membership required to erase the non–monotonicity in the invasion threshold according to Eq.~(\ref{eq:kc}).}
    \label{fig:SM_Fig2}
\end{figure*}

One of the case explored in the main text is defined by $g(k)=\delta_{k,1}$, where individuals participate in only one group at a time. However, modern social settings often involve digital communication, allowing individuals to receive multiple stimuli in parallel. In our model, this corresponds to $\langle k\rangle>1$, and Supplementary Fig. \ref{fig:SM_Fig2} illustrates how this reshapes the phase portrait for fixed group size $p(n)=\delta_{n,3}$ and $p(n)=\delta_{n,5}$.
\medskip

For nonlinear contagion, increasing $k$ progressively alters the influence of temporality. When $g(k)=\delta_{k,1}$ (see Fig. 2 and Supplementary Fig. \ref{fig:SM_Fig2}a), the invasion threshold $\lambda_c(\omega)$ exhibits the non-monotonic behavior: it decreases with $\omega$, reaches a finite minimum, and then increases again. When some individuals can participate in two groups (Supplementary Fig. ~\ref{fig:SM_Fig2}b), this minimum shifts toward $\omega=0$. For $g(k)=\delta_{k,3}$ (Supplementary Fig. ~\ref{fig:SM_Fig2}c), the minimum disappears for sufficiently large nonlinearities, and the threshold becomes strictly increasing in $\omega$: temporal reshuffling now consistently hinders adoption by disrupting the stability needed for reinforcement to accumulate. The same behavior is observed in Supplementary Fig. ~\ref{fig:SM_Fig2}d-f for $p(n)=\delta_{n,5}$.
\medskip

This qualitative change is can be captured analytically by the critical membership, since the case where the non–monotonicity disappears corresponds to
$\omega^\star(k_c)=0$. Imposing this condition in
Eq. (\ref{eq:omega_opt}) gives
\[
2^{\nu/2+1} + 2^{\nu+1} - 4 - k_c(\nu)\,(2^{\nu}-4) = 0,
\]
so that
\begin{equation}
k_c(\nu)
=
\frac{2^{\nu/2+1} + 2^{\nu+1} - 4}{2^{\nu}-4},
\label{eq:kc}
\end{equation}
and its shown in the inset of Supplementary Fig. ~\ref{fig:SM_Fig2}a. 
For $k < k_c(\nu)$, the threshold given by Eq. (\ref{eq:lambdac_general_k}) remains non-monotonic 
and displays an optimal turnover rate. For $k > k_c(\nu)$, $\lambda_c(\omega,k)$ becomes 
strictly increasing: temporality always suppresses the onset of contagion when starting from a susceptible population. Since $k_c(\nu)$ approaches a horizontal 
asymptote in $k=2$, even modest parallel exposure is enough to reverse the role of group switching when synergy is strong.
\medskip

\subsection{Effect of group size}

The comparison between top ($p(n)=\delta_{n,3}$) and bottom ($p(n)=\delta_{n,5}$) rows of Supplementary Fig.~\ref{fig:SM_Fig2} shows how group size shapes: (\textit{i}) the onset of non–monotonicity, i.e., the appearance of a minimum in $\lambda_c(\omega)$; and (\textit{ii}) the value of synergy exponent $\nu$ at which that minimum is pushed to negative group switching rates, $\omega^\star<0$, making the curve monotonic again for all physically allowed $\omega>0$. 
\medskip

For $p(n)=\delta_{n,3}$ and low $k$, the invasion curve remains non–monotonic for all synergy
values $\nu\geq2$, with its
minimum being at $\omega^\star>0$. By contrast, for $p(n)=\delta_{n,5}$ the minimum appears for lower values of $\nu$, and rapidly moves to $\omega^\star<0$, so that
$\lambda_c(\omega)$ becomes strictly decreasing for all $\omega>0$. Therefore, $n$ also reduces the relevance of temporality in determining the order of the transition.

To explain the effect of $n$ in the monotonicity of the invasion threshold curve, we simplify Eq. (\ref{eq:Fn-eq-k}) for a single membership ($g(k)=\delta_{k,1}$) and fixed group size
$p(n)=\delta_{n,n_0}$. Then, the invasion threshold satisfies
\begin{equation}
F_n(x)=\frac{\mu}{\omega},
\qquad
x\equiv\frac{\lambda_c}{\mu+\omega},
\end{equation}
with
\begin{equation}
F_n(x)
=
n\sum_{i=1}^{n-1}
\binom{n-1}{i}(i!)^\nu x^i.
\end{equation}

The invasion threshold becomes non–monotonic when $\lambda_c(\omega)$
develops a minimum. This occurs when
\begin{equation}
F_n'(x^\star)=0,
\qquad
F_n(x^\star)=\frac{\mu}{\omega^\star}.
\end{equation}
The condition $F_n'(x^\star)=0$ selects the extremum, and the second
relation determines its position in $(\lambda,\omega)$–space. Because the combinatorial prefactors
$n\binom{n-1}{i}(i!)^\nu$ grow rapidly with $n$ for all $i\ge2$, the
higher-order terms of $F_n$ and $F_n'$ dominate much earlier for larger
$n$. This produces two distinct consequences:
On the one hand, there is an earlier onset of non-monotonicity: a positive solution $x^\star>0$ to $F_n'(x)=0$ appears for substantially smaller $\nu$ when the group size increases (explaining why the curves for $n=5$ bend at lower synergy than for $n=3$). On the other hand, the minimum of $\lambda_c(\omega)$ is shifted towards negative values of the group switching rate. Once a positive extreme $x^\star$ exists, the corresponding $\omega^\star=\mu/F_n(x^\star)$ decreases sharply with $n$.

Explicitely, for $p(n)=\delta_{n,3}$, the polynomial
\[
F_3(x)=3\,[2x+2^\nu x^2],\qquad
F_3'(x)=3\,[2+2^{\nu+1}x],
\]
admits no positive stationary point, since $F_3'(x)=0$ yields
$x^\star=-2^{-\nu}$, which lies outside the physical domain.
Consequently, $F_3(x^\star)$ never attains values large enough to drive
$\omega^\star=\mu/F_3(x^\star)$ negative. The minimum of
$\lambda_c(\omega)$ therefore always occurs at $\omega^\star>0$, and the
curve remains non–monotonic for all $\nu$.

In contrast, for $p(n)=\delta_{n,5}$,
\[
F_5(x)=5\!\left[\binom{4}{1}x+\binom{4}{2}2^\nu x^2
+\binom{4}{3}6^\nu x^3+\binom{4}{4}24^\nu x^4\right],
\]
where the cubic and quartic contributions grow rapidly with $\nu$.
As a consequence, $F_5'(x)=0$ admits a positive solution $x^\star>0$
already for moderate synergy. Moreover, the corresponding value
$F_5(x^\star)$ increases sharply with $\nu$, so that
$\omega^\star=\mu/F_5(x^\star)$ can become negative. When this occurs,
the minimum is pushed outside the physical domain $\omega>0$, and the
invasion threshold becomes strictly decreasing, as observed in
the bottom row of Supplementary Fig.~\ref{fig:SM_Fig2}.
\newpage

\newpage

\section{Supplementary Note 5: Persistence threshold}

We derive the analytical expression of the persistence threshold
$\lambda_{\mathrm{p}}(\omega,\mu)$ for a hypergraph with three-body
interactions ($p(n)=\delta_{n,3}$) under the single-membership condition ($g(k)=\delta_{k,1}$),
with an explicit recovery rate $\mu$.

\subsection*{Stationary relations}

Let $f_i$ denote the stationary probability that a triplet contains
$i=0,1,2,3$ infected individuals. The detailed-balance relations between
consecutive occupancy states are
\begin{equation}
(i+1)\,\alpha\, f_{i+1}
=
(3-i)\,(\beta_i + \gamma)\, f_i,
\qquad i=0,1,2,
\label{eq:ame-detailed-balance-mu}
\end{equation}
where $\alpha = \mu + \omega(1-I)$, $\gamma = \omega I$. For single membership,
the stationary node equation $s_1=\mu/(\mu+r)$ gives $I = 1 - s_1 = \frac{r}{\mu+r}$. Inside each triplet the synergistic infection rates are
\[
\beta_0=0,
\qquad
\beta_1=\lambda,
\qquad
\beta_2=2^{\nu}\lambda.
\]
Iterating~\eqref{eq:ame-detailed-balance-mu}, all $f_i$ can be written in
terms of $f_0$:
\begin{equation}
f_1 = \frac{3\gamma}{\alpha} f_0,
\qquad
f_2 = \frac{\lambda + \gamma}{\alpha} f_1,
\qquad
f_3 = \frac{2^{\nu}\lambda + \gamma}{3\alpha} f_2.
\label{eq:fi-mu}
\end{equation}

The effective infection pressure $r$ reads
\begin{equation}
r
=\mathcal{M}[I(r)]=
\frac{\displaystyle\sum_{i=0}^3 (3-i)\,\beta_i f_i}
     {\displaystyle\sum_{i=0}^3 (3-i)\, f_i}.
\label{eq:r-def-mu}
\end{equation}
Substituting~\eqref{eq:fi-mu} into~\eqref{eq:r-def-mu}, one finds that
$r$ satisfies the quadratic equation
\begin{equation}
a_2(\lambda,\omega,\mu)\,r^2
+
a_1(\lambda,\omega,\mu)\,r
+
a_0(\lambda,\omega,\mu)
=0,
\label{eq:quad-r-mu}
\end{equation}
where, up to an overall common prefactor,
\begin{align}
a_2(\lambda,\omega,\mu)
&= (\mu+\omega)^2 + \lambda\,\omega,
\\[1mm]
a_1(\lambda,\omega,\mu)
&= -2^{\nu}\lambda^2\omega
   - 2^{\nu}\lambda\,\omega^2
   - \lambda\,\mu\,\omega
   + 2\mu(\mu+\omega)^2,
\\[1mm]
a_0(\lambda,\omega,\mu)
&= -\Bigl(
     2^{\nu}\lambda^2\mu\,\omega
   + 2\lambda \mu^2\omega
   + 2\lambda \mu \omega^2
   - \mu^2(\mu+\omega)^2
   \Bigr).
\end{align}
Equation~\eqref{eq:quad-r-mu} describes the non-zero stationary states of
the dynamics; the absorbing branch $r=0$ has been factored out.

\subsection*{Closed expression of the persistence threshold}

For fixed $(\omega,\mu)$, the boundary of bistability corresponds to a
saddle-node bifurcation of~\eqref{eq:quad-r-mu}, i.e.\ to
\begin{equation}
\Delta_r(\lambda,\omega,\mu)
=
a_1(\lambda,\omega,\mu)^2
-
4\,a_2(\lambda,\omega,\mu)\,a_0(\lambda,\omega,\mu)
=0.
\label{eq:discriminant-mu}
\end{equation}
Expanding~\eqref{eq:discriminant-mu} in powers of $\lambda$ yields a
quadratic equation of the form
\begin{equation}
A_{\mathrm{p}}(\omega,\mu)\,\lambda^2
+
B_{\mathrm{p}}(\omega,\mu)\,\lambda
+
C_{\mathrm{p}}(\omega,\mu)
=0,
\label{eq:lambda-p-eq-mu}
\end{equation}
with coefficients
\begin{align}
A_{\mathrm{p}}(\omega,\mu)
&=
2^{\nu+1}\bigl(2^{\nu}\omega + 3\mu\bigr),
\\[1mm]
B_{\mathrm{p}}(\omega,\mu)
&=
4^{\nu}\omega^{2}
+ 2^{\nu+1}\mu\,\omega
+ 8\mu\,\omega
+ 9\mu^{2},
\\[1mm]
C_{\mathrm{p}}(\omega,\mu)
&=
-4\bigl(2^{\nu}-2\bigr)(\omega+\mu)^{2}.
\end{align}
The persistence threshold is determined by the positive solution of
Eq. \eqref{eq:lambda-p-eq-mu}:
\begin{equation}
\lambda_{\mathrm{p}}(\omega,\mu)
=
\frac{
 -B_{\mathrm{p}}(\omega,\mu)
 + \sqrt{
    B_{\mathrm{p}}(\omega,\mu)^{2}
    - 4 A_{\mathrm{p}}(\omega,\mu) C_{\mathrm{p}}(\omega,\mu)
   }
}{
 2 A_{\mathrm{p}}(\omega,\mu)
}.
\label{eq:lambda-p-final-mu}
\end{equation}

Note that the persistence threshold in Eq.~\eqref{eq:lambda-p-final-mu}
is obtained by projecting the stationary AME dynamics onto the effective
infection pressure $r$ and enforcing the saddle–node condition on the
resulting quadratic equation.
This reduction only becomes asymptotically exact in the annealed limit $\omega \rightarrow \infty$. The
exact bistability boundary over the entire parameter range can instead
be determined directly from the stationary self-consistency condition
in Eq.~(11) of the main text, by imposing the saddle–node condition on
the full stationary AME system.

\subsubsection*{Annealed limit}

The asymptotic behaviour of the persistence threshold for
$\omega \to \infty$ follows directly from the discriminant condition Eq.
 \eqref{eq:discriminant-mu}, using the explicit coefficients
$A_{\mathrm p}(\omega,\mu)$, $B_{\mathrm p}(\omega,\mu)$ and
$C_{\mathrm p}(\omega,\mu)$ defined above.
Since these coefficients are at most quadratic in $\omega$,
the discriminant $\Delta_r(\lambda,\omega,\mu)
=
a_1^2 - 4 a_2 a_0$ is a quartic polynomial in $\omega$.
Expanding Eq. \eqref{eq:discriminant-mu} for large $\omega$ and retaining
the leading contribution, we obtain
\[
\Delta_r(\lambda,\omega,\mu)
=
\omega^4\,
\lambda\Bigl(
4^{\nu}\lambda
+ 8\mu
- 2^{\nu+2}\mu
\Bigr)
+ \mathcal{O}(\omega^3).
\]
Imposing $\Delta_r=0$ in the highly–annealed limit and discarding the
trivial solution $\lambda=0$, we obtain
\begin{equation}
\lim_{\omega \to \infty} \lambda_p(\omega,\mu)
=
\mu\,
\frac{4(2^{\nu}-2)}{4^{\nu}}
=
\mu\,
\frac{4(2^{\nu}-2)}{2^{2\nu}},
\label{eq:lp_lim}
\end{equation}
and therefore, for large $\omega$ the persistence threshold approaches a
horizontal asymptote, as shown in Fig.~2d of the main text.

\subsubsection*{Annealed limit}

The asymptotic behaviour of the persistence threshold when $\omega\to\infty$ can be obtained
directly from the discriminant condition
$\Delta_r(\lambda,\omega,\mu)=0$. Since $a_2$, $a_1$ and $a_0$ are at
most quadratic in $\omega$, the discriminant
\[
\Delta_r(\lambda,\omega,\mu)
=
a_1(\lambda,\omega,\mu)^2
-
4\,a_2(\lambda,\omega,\mu)\,a_0(\lambda,\omega,\mu)
\]
is a quartic polynomial in $\omega$. Expanding in powers of $\omega$ and
keeping the leading term, one finds
\begin{equation}
\Delta_r(\lambda,\omega,\mu)
=
\omega^4\,
\lambda\Bigl(
4^{\nu}\lambda
+ 8\mu
- 2^{\nu+2}\mu
\Bigr)
+ \mathcal{O}(\omega^3),
\end{equation}
where we used $4^{\nu}=2^{2\nu}$. For the discriminant to vanish in the
highly--annealed limit, the coefficient of $\omega^4$ must be zero,
which (discarding the trivial solution $\lambda=0$) yields
\begin{equation}
4^{\nu}\lambda
+ 8\mu
- 2^{\nu+2}\mu
= 0.
\end{equation}
Solving for $\lambda$ we obtain
\begin{equation}
\lim_{\omega \to \infty} \lambda_p(\omega,\mu)
=
\mu\,
\frac{4\bigl(2^{\nu}-2\bigr)}{4^{\nu}}
=
\mu\,
\frac{4\bigl(2^{\nu}-2\bigr)}{2^{2\nu}}.
\label{eq:lp_lim}
\end{equation}
Therefore, the persistence line approaches, for large $\omega$, the
horizontal asymptote, as shown in Fig.~2d of the main text.

\subsection{Width of the bistable region}

We can obtain a closed expression for the width of the bistability area in the mean field limit by subtracting Eqs. (\ref{eq:lc_lim}) and (\ref{eq:lp_lim}). It reads
\[
\lim_{\omega \to \infty} \Delta\lambda(\omega)
= \mu\left(\frac{1}{2}-\frac{4(2^{\nu}-2)}{2^{2\nu}}\right)=\frac{\mu}{2}\left(1-2^{\,2-\nu}\right)^{\!2}.
\]

\newpage

\section{Supplementary Note 6: Real-world structures}

This Supplementary Note details the analysis of seven face-to-face proximity datasets collected in different social environments: a primary school~\cite{stehle2011high}, a conference~\cite{isella2011s}, a hospital~\cite{vanhems2013estimating}, a village in Malawi~\cite{ozella2021using}, two editions of a workplace study~\cite{genois2018can}, and a high school~\cite{mastrandrea2015contact} (see Table \ref{table:t1}). All datasets have finite temporal resolution (20 s), and all contacts recorded within the same window are considered simultaneous. Below, we describe the preprocessing pipeline, the extraction of instantaneous group structures, the tracking of groups across time, and the estimation of the effective group switching rate and the empirical distributions $\{p(n)\}$ and $\{g(k)\}$. These quantities form the structural input used in Fig. 3 and Fig. 4 of the main text.

\begin{table}[h]
\footnotesize
\centering
\begin{tabular}{c@{\hskip 14pt}c@{\hskip 14pt}c@{\hskip 14pt}c@{\hskip 14pt}c}
\toprule
\textbf{Dataset} & \textbf{Context} & $\boldsymbol{\langle \tau \rangle} \; (s)$ & $\boldsymbol{\langle \omega \rangle} \; (s^{-1})$ & \textbf{Reference} \\
\midrule
InVS15     & Workplace              & $1.89 \times 10^{2}$ & $5.3 \times 10^{-3}$ & ~\cite{genois2018can} \\
LyonSchool & Primary school         & $1.05 \times 10^{2}$ & $9.5 \times 10^{-3}$ & ~\cite{stehle2011high} \\
LH10       & Hospital               & $1.41 \times 10^{2}$ & $7.1 \times 10^{-3}$ & ~\cite{vanhems2013estimating} \\
Thiers13   & High school            & $2.08 \times 10^{2}$ & $4.8 \times 10^{-3}$ & ~\cite{mastrandrea2015contact} \\
SFHH       & Scientific conference  & $1.30 \times 10^{2}$ & $7.7 \times 10^{-3}$ & ~\cite{isella2011s} \\
Malawi     & Village                & $1.61 \times 10^{2}$ & $6.2 \times 10^{-3}$ & ~\cite{ozella2021using} \\
\bottomrule
\end{tabular}
\caption{\justifying \textbf{Real-world face-to-face interaction datasets}. For each dataset we report its social context, the average residence time $\langle\tau\rangle$, the effective group switching rate $\langle \omega \rangle$ measured from the data, and the corresponding reference.}
\label{table:t1}
\end{table}

\subsection{Preprocessing of real face-to-face proximity data}

Following the protocol of Iacopini et al.~\cite{iacopini2024temporal}, we preprocess the raw interaction record to ensure temporal consistency and remove unreliable detections.
\medskip

\noindent\textit{Data Cleaning}: For each dataset, we ($\textit{i}$) remove invalid interactions where an individual scans itself or the identifiers are missing; (\textit{ii}) perform gap filling so that if an interaction $(i,j)$ is observed at times $t-1$ and $t+1$ but missing at $t$, we impute a contact at $t$; (\textit{iii}) discard transient contacts appearing at a single isolated timestamp; and (\textit{iv}) apply triadic closure; adding a link ($i,k$) whenever at time $t$ we observe ($i,j$) and ($j,k$), to merge fragmented detections.
\medskip

\noindent\textit{Construction of Instantaneous Graphs}: For each discrete timestamp $t$, we construct an undirected graph $G_t = (V_t, E_t)$ where $V_t$ is the set of active individuals during window $t$, and $E_t = \{(i,j)\}$ contains all interactions recorded within that time window.
\medskip

\noindent\textit{Group Extraction}: Groups at time $t$ are defined as maximal cliques of $G_t$, retaining only cliques of size $n\geq2$. Each time slice is described by a collection of sets $\mathcal{G}_t =
\{g_t^{(1)}, g_t^{(2)}, \dots\}$, where each $g_t^{(k)}$ represents a simultaneously interacting group.
\medskip

\subsection{Tracking groups in real-world structures}

Since the groups at each timestamp are anonymous, we assign persistent identifiers to track their evolution across time.
\medskip

\noindent\textit{Group tracking across time}: Let $g_t\equiv g_t^{(k)}$ and $g_{t-1}\equiv g_{t-1}^{(l)}$ denote two group realizations at consecutive timestamps.Group identity is preserved whenever the two groups share their full core membership. Formally, we quantify this through the overlap similarity:
\begin{equation}
\mathrm{sim}(g_t, g_{t-1})
    = \frac{|g_t \cap g_{t-1}|}{\min(|g_t|,|g_{t-1}|)}.
\end{equation}
By construction, $\mathrm{sim}=1$ whenever one group is a superset of the other. We link the pair $(g_t,g_{t-1})$ if
\[
\mathrm{sim}(g_t,g_{t-1}) \geq \theta,\;\;\;\;\; |g_t \cap g_{t-1}|\geq2,
\]
using $\theta=1$ in this work. Thus, group identity is preserved even if peripheral members join or leave, provided that the smallest group of at least two individuals is fully contained in the largest one.
\medskip

All admissible matches with $\mathrm{sim}\ge\theta$ and $|g_t \cap g_{t-1}|\geq2$ are computed, and non-conflicting assignments are selected via a greedy
maximum-similarity rule. Groups at time $t$ with no match receive new identifiers.
\medskip

\noindent\textit{Residence times}: For each individual $i$, we record the intervals of consecutive timestamps during which it remains in a group with the same persistent identity. The residence time is defined as 
\[
\tau_e = t_{\mathrm{end}} - t_{\mathrm{start}},
\]
and the multiset $\{\tau_e\}$ is computed across all nodes and all such intervals. 
Because reporting occurs at $20$s intervals, residence times are converted to seconds by multiplying by  $\Delta t=20$s.

\subsection*{Effective group switching rate}

The effective group switching rate $\langle \omega\rangle$ quantifies the typical frequency with which individuals reshuffle between groups. Following the definition used in the main text (Eq. (8)), we compute
\begin{equation}
\langle \omega \rangle = \frac{1}{\langle \tau \rangle},
\qquad \text{being} \;\;
\langle \tau \rangle = \frac{1}{E} \sum_{e=1}^{E} \tau_e ,
\end{equation}
where $E$ is the total number of observed residence time intervals. Supplementary Table \ref{table:t1} reports measured values of  $\langle\tau\rangle$  and $\langle\omega\rangle$, which vary widely across contexts, spanning almost an order of magnitude.

\subsection{Size and membership distributions}

Given the temporal sequence of group configurations 
$\{ \mathcal{G}_t \}$, we extract the time series of group counts and the empirical distributions $p(n)$ and $g(k)$.
\medskip

\noindent\textit{Time series of group counts}: The raw timestamps may contain gaps. Therefore, we first define the full axis $[t_{\min},t_{\max}]$ and assign an empty group list $\mathcal{G}_t=\emptyset$ to missing 
snapshots. Then, for each snapshot $t$, we compute the number of groups
\[
N(t) = |\mathcal{G}_t|,
\]
which is used for visualization (first column of Supplementary Fig. \ref{fig:SMFig1}).
\medskip

\noindent\textit{Group-size distribution $p(n)$}: For all groups across all snapshots, we collect the set of sizes
\[
\mathcal{S} = \{\, |g_t| : g_t\in \mathcal{G}_t\}.
\]
If $c_n$ denotes the number of occurrences of size $n$ in $\mathcal{S}$, the normalized group-size distribution is
\[
p(n) = \frac{c_n}{\sum_{m} c_m}.
\]
\medskip

\noindent\textit{Membership-per-snapshot distribution $g(k)$}: For each individual $i$ at time $t$, we compute the number of 
groups the individual participates in,
\[
k_i(t) = \bigl|\{\, g_t\in \mathcal{G}_t: i\in g_t \,\}\bigr|.
\]
We collect the values $k_i(t)$ across all snapshots into the multiset $\mathcal{K}$, and let
$d_k$ denote the total number of occurrences of value $k$ in $\mathcal{K}$. Then, the normalized distribution of memberships per snapshot is
\[
g(k) = \frac{d_k}{\sum_{m} d_m}.
\]
\medskip

Both normalized distributions are shown for every dataset in Supplementary Fig.~\ref{fig:SMFig1} and feed the expressions of the tricritical point and invasion threshold to produce Fig. 3 and Fig. 4.

\begin{figure}
    \centering
    \includegraphics[width=1\linewidth]{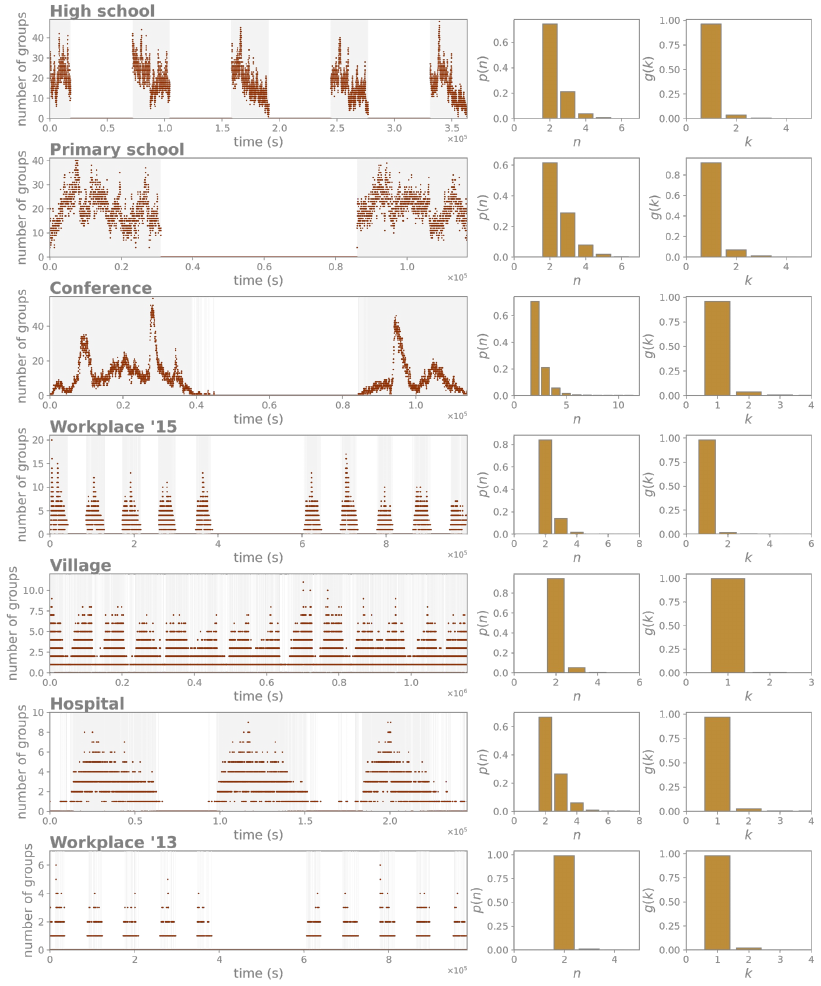}
    \caption{\textbf{Empirical temporal structure of real datasets.} Time series of the number of groups $N(t)$, distribution of group sizes $p(n)$, and distribution of memberships per snapshot $g(k)$. The datasets belong to the different social contexts reported in Table \ref{table:t1}.}
    \label{fig:SMFig1}
\end{figure}

\newpage

\section{Supplementary Note 7: Effect of temporality and heterogeneity in real-world and synthetic structures}

In Supplementary Fig.~\ref{fig:SM_Fig3_1}a we extend Fig.~3a of the main text by representing the tricritical lines for all the empirical face-to-face interaction datasets considered in Table \ref{table:t1}. We show that the qualitative phenomenology reported in Fig.~3a is robust across structures with different levels of size heterogeneity and group switching. While the precise location of the tricritical line shifts from dataset to dataset, its shape remains unchanged. Furthermore, the inset reports the actual numerical values of the effective structural coupling $Q$ (beyond the rank presented in Fig.~4).
\medskip

To further disentangle the role of heterogeneity in the onset of multistability, in Supplementary Fig.~\ref{fig:SM_Fig3_1}b we perform a systematic analysis on synthetic structures with varying group size heterogeneity. We fix the membership distribution to match that of the high-school dataset \cite{mastrandrea2015contact} and vary only the group-size distribution, imposing truncated power-law forms $p(n)\sim n^{-\gamma}$ with cutoff $n_{\text{max}}=6$ (consistent again to the high-school dataset, see Supplementary Fig. \ref{fig:SMFig1}). Tuning the level of heterogeneity through the exponent $\gamma$, we observe that increasing heterogeneity turns the tricritical line from being a monotonous function to a non-monotonous one.
\medskip

Finally, Supplementary Fig.~\ref{fig:SM_Fig3_2} complements Fig. 3b by showing the quenched ($\omega=0$ in  Supplementary Fig.~\ref{fig:SM_Fig3_2}a) and close-to-annealed ($\omega=100$ in  Supplementary Fig.~\ref{fig:SM_Fig3_2}b) phase diagrams. The quenched limit displays continuous transitions with non-monotonous prevalence growth, showcasing different plateaus corresponding to different levels of mesoscale localization. This is consistent with previous findings by St-Onge et al. \cite{st2021master,st2022influential} on the role of heterogeneity on the phase transition. The annealed limit displays discontinuous transitions, equivalent to the phenomenology obtained by Iacopini et al. \cite{iacopini2019simplicial}. This highlights that multistable active phases with multiple coexisting active states are observed in neither the annealed nor the quenched limits.

\begin{figure}[h]
    \centering
    \includegraphics[width=1\linewidth]{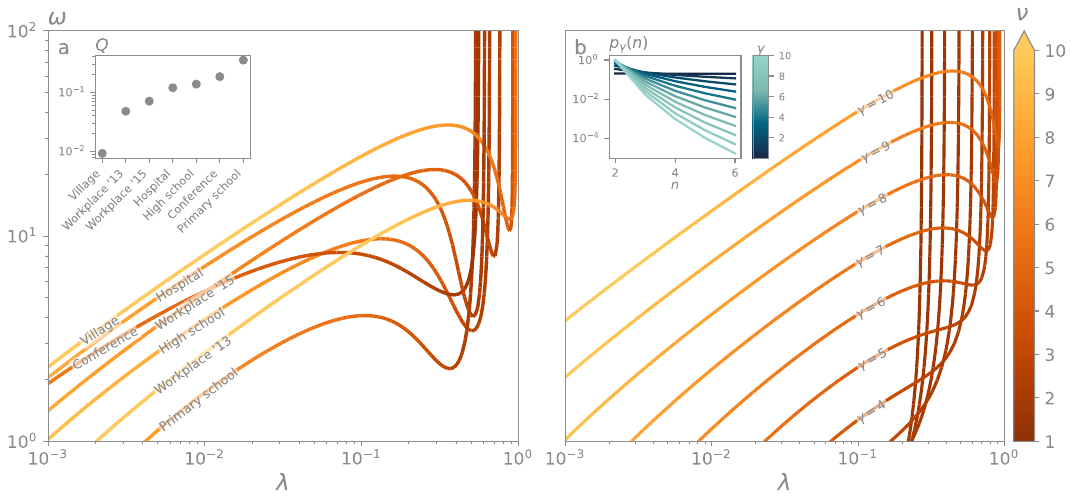}
\caption{
\textbf{Interplay between temporality and size heterogeneity yields multistability.}
\textbf{a} Tricritical lines (solid curves) in the $(\lambda,\omega)$ plane for the empirical face-to-face datasets in Table \ref{table:t1}. The inset reports the effective structural coupling of each dataset according to Eq. (25). \textbf{b} Tricritical lines for synthetic structures with fixed membership distribution (matching the high-school dataset) and power-law group-size distributions $p(n) \sim n^{-\gamma}$ with cutoff $n_{\max}=6$ (reported in the inset). 
}
\label{fig:SM_Fig3_1}
\end{figure}

\begin{figure}[h]
    \centering
    \includegraphics[width=1\linewidth]{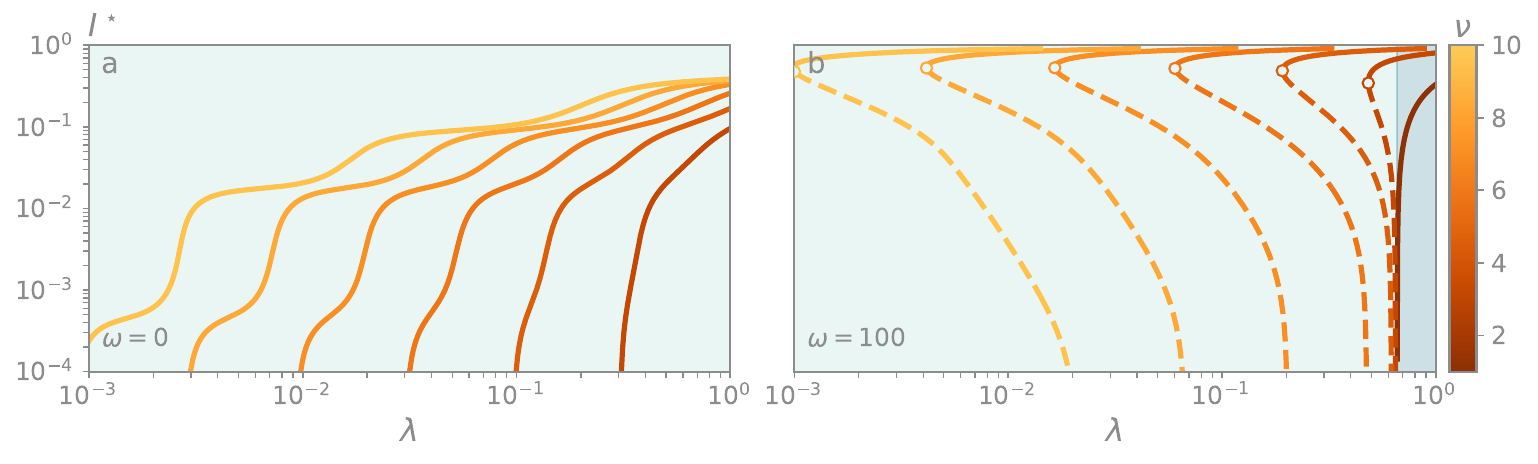}
\caption{
\textbf{Cross-sections of the phase diagram at fixed group switching rates.} Stationary prevalence $I^\star$ as a function of the adoption rate $\lambda$ for increasing values of the synergy exponent $\nu$ (color-coded) in the quneched ($\omega=0$, \textbf{a}) and close-to-annealed ($\omega=100$, \textbf{b}) limits. Solid (dashed) lines denote stable (unstable) stationary states.
}
\label{fig:SM_Fig3_2}
\end{figure}

\newpage
.

\newpage

\section{Supplementary Note 8: Effect of homogenizing temporality in Chowdhary et al.}

In Ref. \cite{chowdhary2021simplicial}, Chowdhary et al. analyse simplicial contagion on higher-order networks, and their Fig. 1 shows that temporality hinders the onset of contagion dynamics, even in the linear case. In Supplementary Fig. \ref{fig:SM_Fig3} we show that this effect does not originate from temporality itself, but from two
exogenous factors introduced by their modelling framework.
\medskip

First, their temporal model rewires the structure at every time step without preserving the degree sequence. As a result, node degrees fluctuate across time and across nodes. This induces artificial structural heterogeneity that does not stem from temporality, but from repeatedly resampling the network from scratch.
\medskip

Second, they use a Markov chain approach (MMCA) that assumes independent contagion pathways. Specifically, the probability
that one node $i$ out of the $N$ nodes in a structure is infected at time $t+1$ is assumed to follow
\begin{equation}
p_i(t+1)
=
\bigl(1-q_i(t)\,q_{i,\Delta}(t)\bigr)\bigl(1-p_i(t)\bigr)
+
(1-\mu)\,p_i(t),
\label{eq:chow_update}
\end{equation}
where $p_i(t)$ is the infection probability of node $i$ at time $t$,
$\mu$ is the recovery probability, $q_i(t)$ is the probability that $i$ is not
infected via pairwise interactions with its neighbours, and
$q_{i,\Delta}(t)$ is the probability that $i$ is not infected via any of
its two-simplices. These are approximated as
\begin{equation}
q_i(t)
=
\prod_{j\in\Gamma_i(t)}
\bigl(1-\beta\,p_j(t)\bigr),
\label{eq:chow_qi}
\end{equation}
\begin{equation}
q_{i,\Delta}(t)
=
\prod_{j,\ell\in\Delta_i(t)}
\bigl(1-\beta_\Delta\,p_j(t)\,p_\ell(t)\bigr),
\label{eq:chow_qDelta}
\end{equation}
where $\Gamma_i(t)$ is the set of edges attached to $i$, at time $t$, and $\Delta_i(t)$ the set
of triangles incident to $i$. The constants $\beta$ and
$\beta_\Delta$ are the infection probabilities associated with edges and triangles, respectively.
\medskip

\begin{figure}[h]
    \centering
    \includegraphics[width=0.75\linewidth]{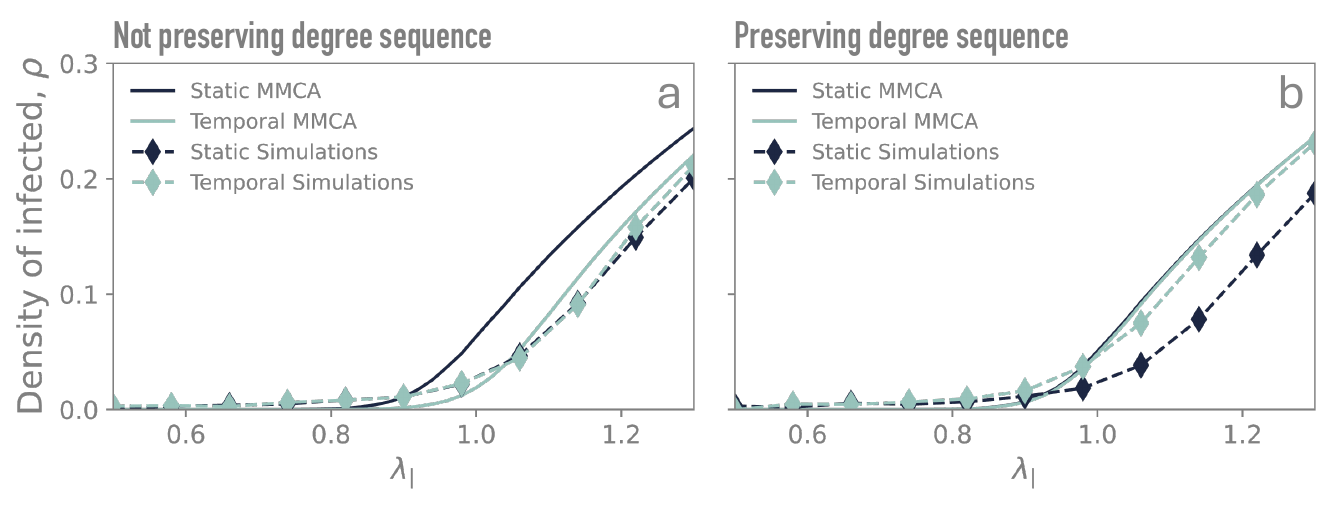}
\caption{\textbf{Effect of preserving the degree sequence on temporal SIS
dynamics.} Panels \textbf{a} and \textbf{b} compare the stationary prevalence
$\rho = N^{-1}\sum_i p_i$ as a function of the infection rate $\lambda=\beta\langle k\rangle/\mu$
for static and temporal networks in the linear case ($\beta_\Delta = 0$).
Temporal networks are generated by randomly rewiring the edges at
each step. In \textbf{a}, the rewiring does not preserve the degree
sequence, so node degrees fluctuate in time, as in the temporal model
of Ref.~\cite{chowdhary2021simplicial}. Under these conditions, the
iteration of the Markov chain equations
(\ref{eq:chow_update})--(\ref{eq:chow_qDelta}) predicts a suppression of
contagion in the temporal setting~\cite{chowdhary2021simplicial}. In \textbf{b}, the rewiring does preserve the degree sequence at each snapshot. In this case, the Markov chain approach yields identical prevalence curves for the static
and temporal networks, demonstrating that the suppression observed in \textbf{a} arises from temporal degree fluctuations rather than temporality itself. Monte Carlo simulations (markers) show the opposite trend: when the degree sequence is preserved, temporality shifts the threshold to lower values of $\lambda$, in agreement with known results on temporal networks~\cite{valdano2015analytical,st2018phase}.
When the degree sequence is not preserved (panel \textbf{a}), this structural heterogeneity partly compensates this effect, but the temporal curve remains above the static one in the supercritical
regime. All structures are Random Simplicial Complexes (RSC, see Ref. \cite{iacopini2019simplicial}), have $N=500$ nodes and average degree $\langle k\rangle=12$. We consider the linear case $\beta_\Delta=0$, and the recovery rate is set to $\mu=0.1$, and the markers of the Monte Carlo simulations represent the average of 1000 realizations.}

    \label{fig:SM_Fig3}
\end{figure}

Supplementary Fig. \ref{fig:SM_Fig3}a, compares the stationary state $\rho=N^{-1}\sum_ip_i$ obtained from the iteration of Eqs. (\ref{eq:chow_update})-(\ref{eq:chow_qi}) on a static structure with the corresponding temporal dynamics when the degree sequence is not preserved. In the linear case ($\beta_\Delta=0$), the result reproduces the qualitative trend reported in Ref.~\cite{chowdhary2021simplicial}: the temporal structure appears to hinder contagion.
\medskip

However, Supplementary Fig. \ref{fig:SM_Fig3}b shows the same comparison when the temporal evolution preserves the degree sequence. In this case, the Markov chain approximation predicts identical stationary states for the static and temporal networks. This demonstrates that the differences observed in Supplementary Fig. \ref{fig:SM_Fig3}a are fully driven by the temporal fluctuations in the degrees of the nodes, and not by temporality itself.
\medskip

Finally, Supplementary Fig.~\ref{fig:SM_Fig3} also includes stochastic Monte Carlo simulations. When the degree sequence is preserved (Supplementary Fig. \ref{fig:SM_Fig3}b), temporality shifts the epidemic threshold to lower values of $\lambda$, in agreement with the well-established behaviour of temporal networks~\cite{valdano2015analytical,st2018phase}. When the degree sequence is not preserved (Supplementary Fig. \ref{fig:SM_Fig3}a), the imposed degree
fluctuations partially counteract the facilitation effect of temporality, but the temporal prevalence curve still lies above the static one in the
supercritical regime.
\medskip

\end{document}